# Deep Convolutional Autoencoders as Generic Feature Extractors in Seismological Applications

**Qingkai Kong[1], Andrea Chiang[1], Ana C. Aguiar[1], M. Giselle Fernández-Godino[1], Stephen C. Myers[1], Donald D. Lucas[1]**

[1] Lawrence Livermore National Laboratory.

Corresponding author: Qingkai Kong (kong11@llnl.gov)

Present address:

**Highlights:**

- Autoencoders can serve as feature extractors for different applications
- Overcomplete models perform better than undercomplete models in most cases
- Adding one extra CNN layer after the encoder extractor yield better results
- Fine tuning all the designed layers works better than only updating the last layers

**Abstract**

The idea of using a deep autoencoder to encode seismic waveform features and then use them in different seismological applications is appealing. In this paper, we designed tests to evaluate this idea of using autoencoders as feature extractors for different seismological applications, such as event discrimination (i.e., earthquake vs. noise waveforms, earthquake vs. explosion waveforms), and phase picking. These tests involve training an autoencoder, either undercomplete or overcomplete, on a large amount of earthquake waveforms, and then using the trained encoder as a feature extractor with subsequent application layers (either a fully connected layer, or a convolutional layer plus a fully connected layer) to make the decision. By comparing the performance of these newly designed models against the baseline models trained from scratch, we conclude that the autoencoder feature extractor approach may only outperform the baseline under certain conditions, such as when the target problems require features that are similar to the autoencoder encoded features, when a relatively small amount of training data is available, and when certain model structures and training strategies are utilized. The model structure that works best in all these tests is an overcomplete autoencoder with a convolutional layer and a fully connected layer to make the estimation.

# 1 Introduction

Machine learning models, especially recently developed deep learning models, have the capability to extract features, which are measurable properties or characteristics of the studied phenomenon, from images, texts, time series and many other types of data (Goodfellow et al., 2016; LeCun et al., 2015). Many good reviews are available that describe deep learning applications in seismology and the broad geosciences (Bergen et al., 2019; Karpatne et al., 2019; Kong et al., 2019; Lary et al., 2016). In seismology specifically, great performance has been achieved in event detection and discrimination (Kong et al., 2016; Li et al., 2018; Linville et al., 2019; Meier et al., 2019; Perol et al., 2018), seismic phase picking (Mousavi et al., 2020; Ross et al., 2018; Zhou et al., 2019; Zhu & Beroza, 2019), denoising (Chen et al., 2019; Saad & Chen, 2020; Tibi et al., 2021; Zhu et al., 2019), and lab experiment predictions (Rouet-Leduc et al., 2017). To highlight a few applications related to the work presented here, Ross et al. (2018) and Zhu & Beroza (2019) developed deep learning based approaches for phase picking, which are now adopted widely to estimate the P and S arrivals (Chai et al., 2020; Graham et al., 2020; Park et al., 2020; Wang, Schmandt, Zhang, et al., 2020). They designed deep learning models that automatically extract the waveform characteristics distinguishing the P phase, S phase and noise to make decisions about P and S arrivals on the seismic waveform. After training with large amounts of seismic data, the two models generalize well with new input data. Linville et al. (2019) explored using convolutional and recurrent neural networks to discriminate  explosive and tectonic sources at local distances, they showed that developed models can successfully determine the source type of the events at an accuray above 99%.

An autoencoder is a machine learning model that can be used to learn efficient representations (encoding) from a set of data, and then recover the data from these encoded representations. Deep autoencoders have been used in many different applications, such as compression, denoising, dimensionality reduction, and feature extraction (Baldi, 2012; Liu et al., 2017). Particularly, using autoencoders to extract features for different tasks show great promise (Ditthapron et al., 2019; Gogna & Majumdar, 2019; Kunang et al., 2018; Xing et al., 2015). In seismology, applications use autoencoders to extract compressed feature representations for different uses, such as clustering and regression,. (Bianco et al., 2020; Jenkins et al., 2021; Mousavi, Zhu, et al., 2019; Snover et al., 2021; Spurio Mancini et al., 2021). For example, Snover et al. (2021) use convolutional autoencoders to learn latent features from the spectrograms of the ambient seismic noise data from the Long Beach dense seismic array, and then use deep clustering algorithms to identify the source of the different clusters. Mousavi, et al. (2019) use the features extracted from the autoencoders to discriminate local waveforms from teleseismic waveforms and determine first motion polarities. Inspired by these applications, we test whether encoded features from autoencoders may subsequently be used for applications of seismological interest, such as event discrimination and phase picking, among other applications. Taking this approach has the following benefits compared to training the deep learning models from scratch: first, extracting features in this way would streamline the processing pipeline and improve the usage of these deep learning models, because we only need one big generic waveform dataset for the autoencoder model that learns the waveform characteristics in an unsupervised way (because it only reconstructs the waveforms). Second, after training, the model can be used in many different applications where labeled training data are sparse. We only need a small dataset for fine tuning to accommodate the new use cases. In a sense, this is a special case of transfer learning (Bengio, 2012; Shin et al., 2016; Tan et al., 2018), where we train a deep neural network model on a problem with large amounts of data expecting that the extracted features will be transferable to other tasks by fine tuning of newly added layers or previously learned layers. In this case, we use the encoder portion of the deep convolutional autoencoder to transfer the learned features to different problems.

In order to test the effectiveness of extracting and transferring seismic features, we systematically evaluate this method using different datasets for three different seismological applications: noise vs. earthquake classification, P wave arrival picking, and explosion vs. earthquake discrimination. We tested the use of overcomplete and undercomplete autoencoders, using a different number of feature maps in the main encoder layers, adding an extra convolutional layer before the fully connected layer to make the decision, and training with different approaches to evaluate the performance. By comparing the performance of these newly designed models against the baseline models trained from scratch, we conclude that the autoencoder feature extractor approach may have advantages in certain circumstances: such as when the target problems require features that are similar to the autoencoder features, when a relatively small amount of training data is available for the target problem, and when certain model structures and training strategies are utilized. The model structure that works best in all these tests is an overcomplete autoencoder with a convolutional layer and a fully connected layer to make the estimation. In almost all the other cases, a model trained specifically for the problem (a.k.a. baseline model) outperforms the autoencoder-based approaches.

## 2 Methods

### 2.1 Overview

The idea behind this paper is to first train an autoencoder on a large number of seismic waveforms to encode features then reconstruct the signals. The trained encoder should capture the main characteristics of the seismic waveforms, and thus can work as a feature extractor. By concatenating more convolutional layers and/or fully connected layers (so called application layers) to the encoded layer, the combined model can be used in different applications. Figure 1 shows the whole workflow of the method. The top big blue solid boxes in Figure 1 illustrate the autoencoder structures that are utilized here. After training, the encoder portion of the autoencoder (i.e., the layers from the input to the bottleneck layer, blue dotted box) is cut out and appended to the application layers, which contain an optional convolutional layer, a fully connected layer, and a decision layer using a sigmoid or rectified linear unit (ReLU) for making decisions. To be able to apply this approach to different applications, we used two different training approaches. In the first approach, only the application layers at the end were trained, with all the encoder layers locked. In the second approach, both the application layers as well as the encoder layers were tuned with a much smaller learning rate than originally used. The following sections will explain these training approaches in more detail.

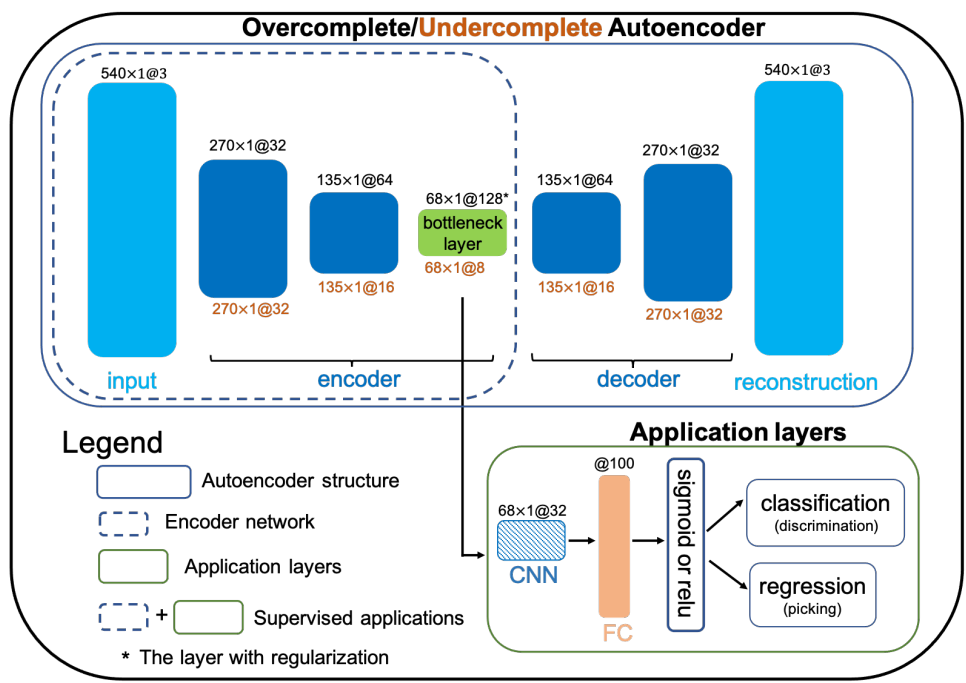

**Figure 1.** The workflow of the experiments. The above solid blue boxes show the structure of the designed autoencoder. Black text labels on the top of the layers represent the overcomplete autoencoder, orange color text labels on the bottom represent the undercomplete autoencoder, and the dotted blue box around the input and bottleneck layers contains the encoder. The format of the text in m×1@n indicates the feature map (or input) is m pixels wide and 1 pixel in height, with n channels (or the number of feature maps). The encoder output (green block) contains the learned features from the so-called bottleneck layer. These learned features are fed into the application layers containing an optional convolutional network (CNN) that provides another layer of feature extraction (tested with and without), a fully connected (FC) layer with 100 neurons is applied to the flattened features, and a sigmoid or ReLU activation function for different applications.

## 2.2 Autoencoders

An autoencoder is a neural network that is trained to attempt to copy its input to its output (Goodfellow et al., 2016). It generally comprises two parts, the encoder and decoder. The encoder frequently contains a series of layers to extract features of the input, and passes these features to another series of layers, the decoder, to reconstruct the input. To make autoencoders useful at learning features and not simply copying the inputs to the outputs, we can follow two paths. One path is to constrain the last encoder layer to have smaller total dimensions than the input dimensions, essentially making a bottleneck in the middle of the whole autoencoder. For example, in Figure 1, our input dimension is 1620 (540×3), while the bottleneck layer dimensions are 544 (68×8), which essentially compresses the data to about one third of the input dimensions. This type of autoencoder is called an undercomplete autoencoder, since the bottleneck layer is smaller than the input dimension. Squeezing the dimensions in this way forces the autoencoder to capture the most useful features of the training data. Another path is to extend the bottleneck layer of the encoder to have more dimensions than the input, in our case, 1620 input versus 8704 (68×128) output dimensions, but adding a regularization term to force the bottleneck layer to be sparse, thus constraining the autoencoder to learn useful features instead of simply copy the input. We use a L1 regularization term (10e-5) at the bottleneck layer. Contrastingly, overcomplete autoencoders have been developed because they have greater robustness in the presence of noise and have greater flexibility in learning useful features from the data (Goodfellow et al., 2016).

In our models, the encoder layers use 2D convolutional operations with a kernel size of (3, 1) and strides of (2, 1) to shrink the size of the feature maps to half in the first axis. The decoder upscales the size of the feature maps using transpose 2D convolutional layers with a kernel size of (3, 1) and strides of (2, 1) to reconstruct the waveforms to 540×3.

To train the model, we utilized the Adam optimizer (Kingma, 2015) with mean absolute error as the loss function and implemented an early stopping criterion to avoid overfitting: training stops if performance did not improve over 20 epochs.

## 2.3 Application Layers

After training, the encoder (dotted blue box in Figure 1) can be used as a general feature extractor. By combining the encoder with application layers, we can test whether extracted features are useful for different problems. Essentially, the structure of the model used here is similar to some of the existing models that used the standard convolutional neural network with shrinking dimensions with deeper layers (Ross et al., 2018). The differences between them are the training processes, with the model we designed here using a training approach like transfer learning. We tested two architectures of the application layers: The first one has a single fully connected layer where 100 neurons were added to the encoder and the flattened output of the encoder serves as the input. The second one has a convolutional layer with 32 kernels before the fully connected layer. If we do not use this CNN layer, the features extracted directly from the encoder will be used for making decisions. With the CNN layer added, it provides another mechanism to update and extract the features to make it more adaptable to the new problems, thus achieving better results as will be shown in the results section. A sigmoid or ReLU activation function was used as the output depending on whether it is a classification or regression problem.

## 2.4 Baseline model

A baseline model refers to a simple or existing model that can serve as a performance reference. In this case, we use a model with the same structure as the encoder and application layers, but trained from scratch instead of relying on pre-trained encoder, as our baseline model.

## 2.5 Training setup

Two training schemas were tested. (1) train only the application layers, so the encoder parameters are locked. This is similar to the standard transfer learning approach (Tan et al., 2018), or a special case of transfer learning that only tunes the last fully connected layers. This approach assumes that the locked layers trained on a large amount of data are considered as good feature extractors. In this case, tuning the last layer will help the model accommodate the new patterns in the new dataset that the model will be applied on. This approach usually works well when the features extracted from the locked layers are similar to those in the new dataset, and it works best if the problems are similar or the same. (2) After tuning the application layers, we also fine tune the parameters of the encoder layers (i.e. unlock them), using a very small learning rate. The features trained on a different dataset or task may be so different from the ones in the new problem that fine-tuning these pre-trained layers with a very small initial learning rate can help improve the model performance. Some of the training parameters are shown in Table 1.

| Problem | Training | Monitor metrics | Monitoring Mode | Learning rate | Optimizer and Loss |
|---------|----------|-----------------|-----------------|---------------|--------------------|
| Noise vs EQ  LEN_DB | baseline | Validation accuracy | max | 0.01 | Adam optimizer  Sparse Categorical Crossentropy |
|  | schema 1 |  |  | 0.01 |  |
|  | schema 2 |  |  | 0.00005 |  |
| Explosion vs EQ  Multiple | baseline | Validation accuracy | max | 0.01 | Adam optimizer  Sparse Categorical Crossentropy |
|  | schema 1 |  |  | 0.01 |  |
|  | schema 2 |  |  | 0.00005 |  |
| P phase picking  STEAD | baseline | Validation loss | min | 0.01 | Adam optimizer  Mean Absolute Error |
|  | schema 1 |  |  | 0.01 |  |
|  | schema 2 |  |  | 0.00005 |  |

**Table 1** Training parameters of the models with Adam optimizer. We only save the best model based on the monitored metrics with the corresponding mode, the batch size used is n/100, which is the 1% number of training samples. Validation dataset is 20% of the total training data, shuffle was used in each Epoch. Early stopping was used as a regularization to avoid overfitting, with 20 epochs as the patience parameter, which means if the monitored metric does not improve for 20 epochs, we stop training. Training baseline is the model training from scratch, schema 1 is the model only fine tune the last layers, while schema 2 includes fine tuning encoder layers.

## 3 Data

To test the three different applications, as well as building the autoencoder, we used data from multiple sources. In this section we include a detailed description of the data used in the applications. Example waveforms are shown in Figure 2.

### 3.1 Autoencoder

The data for training the autoencoder comes from the STEAD dataset (Mousavi et al., 2019), which contains about ~1.2 million local earthquake waveforms (with P and S arrival labels). We only used the earthquake waveforms to train the autoencoders. The earthquake waveforms were resampled to 20 samples per second for a total length of 540 (27s) and amplitude normalized from -1 to 1. 576434, 144109, and 308805 waveforms were used for training, validation and testing purposes, respectively. The magnitude, distance and depth distribution are shown in Figures S17-S19 (the blue bars), for more details about the dataset, please refer to Mousavi et al., (2019).

## 3.2 Noise vs. Earthquake

For this problem, we take advantage of the LEN_DB dataset (Magrini et al., 2020), which contains 629,095 three-component earthquake waveforms generated by 304,878 local events and 615,847 noise waveforms. The noise records in the LEN_DB dataset were randomly selected from the period when no earthquake events occurred nearby the stations (see Magrini et al., 2020 for the criterion used). Each waveform is sampled at 20 samples per second with a total of 540 data points. The magnitude, distance and depth distribution are shown in Figures S20-S22.

## 3.3 Explosion vs. Earthquake

This dataset is assembled from 3 different experiments: SPE (Source Physics Experiment) Phase I (Pyle & Walter, 2019), iMUSH/MSH (Imaging Magma Under St. Helens) (Hansen & Schmandt, 2015), and BASE (The Bighorn Arch Seismic Experiment) (Wang, Schmandt, & Kiser, 2020). Overall, it has 9,728 three-component explosion records and 23,645 earthquake records. The SPE Phase I conducted between 2011 and 2016 are a series of underground chemical high-explosive detonations in saturated granite of various sizes and depths. There were five borehole shots ($M_L$ $1.2 - 2.1$, with explosive yields between about 0.1 and 5 ton TNT equivalent (Snelson et al., 2013). The MSH dataset contains 23 explosive sources ($M_L$ 0.9-2.3) and 91 earthquakes ($M_L$ $1.5 - 3.3$) within 75 km of the volcano during the 2014 – 2016 iMUSH project (Hansen & Schmandt, 2015; Wang, Schmandt, & Kiser, 2020). The explosive events are shallow borehole shots and are distributed relatively uniformly in this region. The BASE experiment (Worthington et al., 2016; Yeck et al., 2014) was conducted in 2010 to image the Bighorn Arch in Wyoming. 21 explosive sources ($M_L$ $0.7 - 1.7$, loads 113-907 kg) and 19 earthquakes ($M_L$ $0.3 - 2.7$) recorded by ~90 broadband stations and ~180 short-period stations are in the dataset. These explosion records have different P/S ratios (Pyle & Walter, 2019) and local/code magnitudes (Koper et al., 2021) than natural earthquakes. To make the data consistent with the 20 samples per second (540 data points), we first low-pass filtered data at 10 Hz and resampled it to 20 samples per second. For these records, we cut the original explosion waveforms ten times and the earthquake records four times with a start time randomly selected from -22.5s to the origin of the earthquake to augment the data. A total of 97,280 explosion records and 94,580 earthquake records were obtained for training purposes. Figures S23-S25 show the magnitude, distance and depth distribution of these records.

## 3.4 P phase picking

From the STEAD dataset (Mousavi, Sheng, et al., 2019), we selected all the earthquake waveforms within 100 km with magnitude larger than M2.0. For each earthquake waveform, we first low-pass filtered data at 10 Hz and resampled it to 20 samples per second. Then we cut a window of 540 data points (because the raw waveform is longer) with the start time randomly selected before the P wave arrival. This process returned 168,859 waveforms for training and testing purposes. Note that we also tested with more data as shown later in the results section. In this test, we used data with magnitude larger than M1.5 within 200 km, which returned 425,552 waveforms. The magnitude, distance and depth distribution are shown in Figures S17-S19 as green and orange bars.

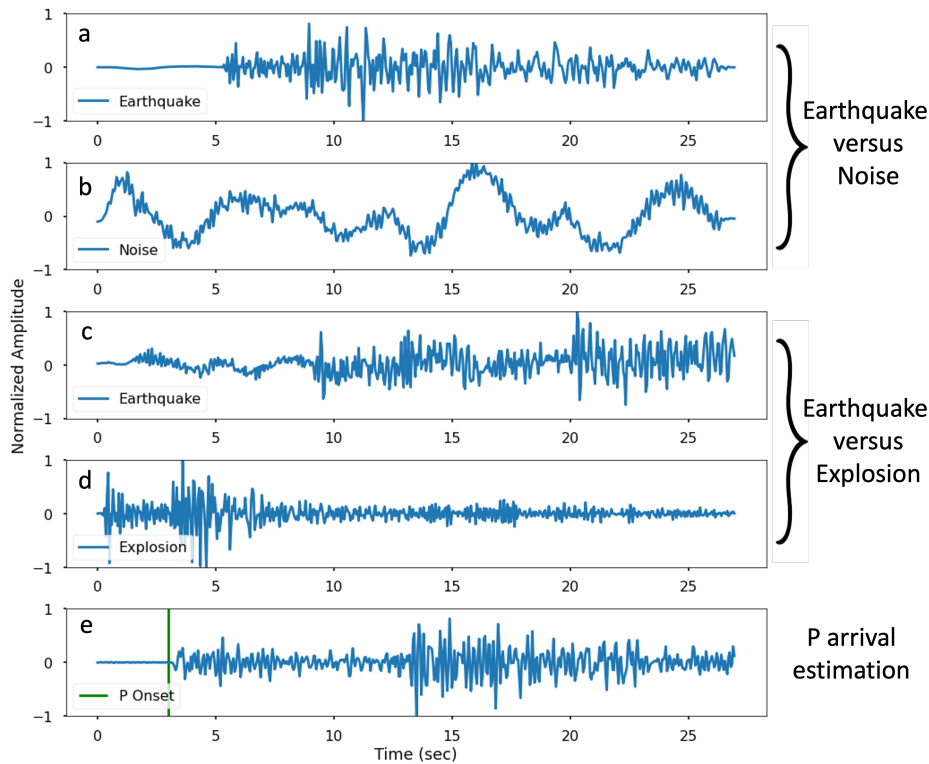

**Figure 2**. Example input waveforms for the three examples. (a) and (b) are used in the earthquake vs. noise problem, where the earthquake waveform in (a) is from a M3.0 earthquake recorded at 36.7 km on station 319A in AE network. (c) and (d) are used in the earthquake vs. explosion problem, where (c) is a waveform from a M2.2 earthquake recorded at 58.7 km on station SWF2 in the MSH experiment and (d) is a waveform from a M1.3 explosion recorded on station SM34 at 22.5 km in the BASE experiment. (e) is a waveform from a M3.7 earthquake recorded by a station 109C in the TA (Transportable Array) network at 75.8 km.

## 4 Results

### 4.1 Autoencoder results

The mean absolute error results for the test data and the different autoencoder architectures are shown in Figure 3. In this figure, we can observe four main results: (1) the performance is getting better when the model has fewer layers (i.e. it has a larger feature map dimension at the bottleneck layer) using the same number of feature maps. Since the bottleneck extracted features are the building blocks of the decoder to reconstruct the waveforms, at the same number of the feature maps, the larger ones are easier for the model to reconstruct. (2) Models with more layers extract smaller features, therefore, we need more feature maps to reconstruct the signal, consistent with point 1. As a result, we see better performance when we increase the number of feature maps (red line is the best performance with lowest errors while the purple line is the poorest performance). (3) The overcomplete models perform better than the undercomplete model, because there are more features that can be extracted in the bottleneck layer. (4) We also notice that when the number of feature maps is smaller than 32, such as the blue and purple lines, the pattern of the lines changed compared to the rest of the models (i.e., the direction of the curvature). We think this is due to the smaller number of feature maps used in the bottleneck layer, which limits the power of combining these extracted features, but this needs further testing to confirm. To compare the performance of the reconstruction of the undercomplete and overcomplete models, Figure 4 shows the outputs of the autoencoders comparing against the inputs. We can see the matching of the waveforms for the overcomplete model is better than those from the undercomplete model, which illustrates the features extracted from the overcomplete model captures more of the characteristics of the waveforms.

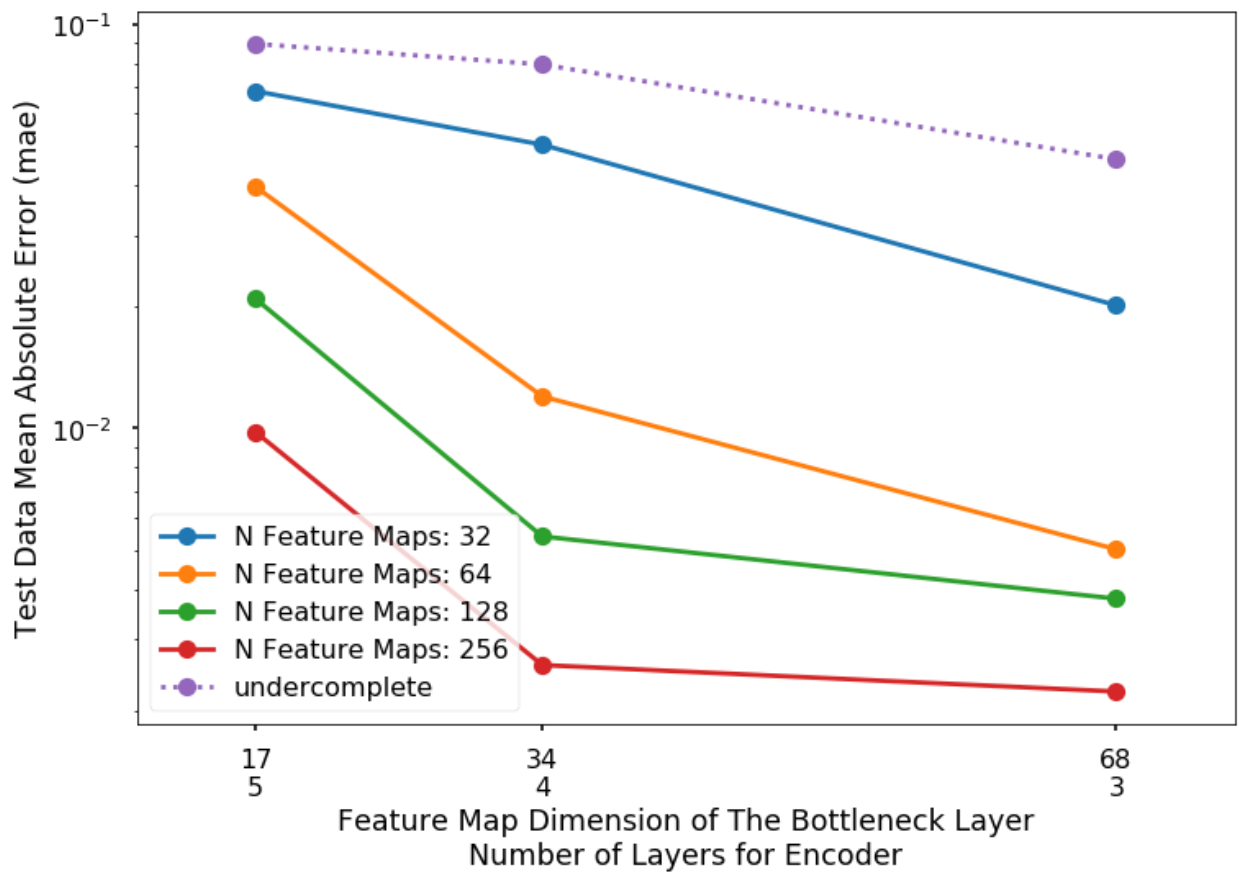

**Figure 3.** The test data performance using mean absolute error for various trained autoencoders. Solid lines are from the overcomplete autoencoders while the dotted line model is from the undercomplete autoencoder. Models with 17, 34, and 68 feature map dimensions in the bottleneck layer were tested. The $2^{nd}$ row in the x axis labels is showing the number of layers used in the encoder. Different colors of the lines represent the number of feature maps used in the $3^{rd}$ and deeper layers.

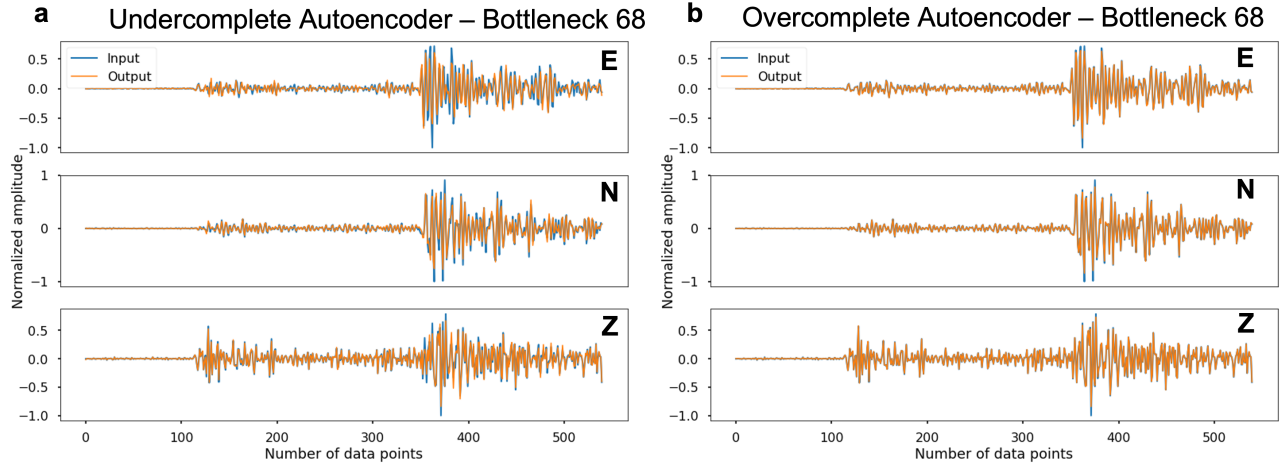

**Figure 4**. Reconstructed waveforms from the autoencoders with bottleneck dimension 68, blue and orange waveforms are the inputs and outputs, respectively. The three rows in each panel are the east-west, north-south and vertical components. (a) outputs from the undercomplete autoencoder. (b) outputs from the overcomplete autoencoder.

## 4.2 Results for Applications

To achieve a good balance between performance and computation cost, we selected 128 feature maps for the undercomplete and overcomplete models for testing. We tested the models when the bottleneck feature map dimensions are 68 and 34 with different amounts of training data. For each model-specific configuration, we ran 5 different training/testing instances varying the initialization of the model weights as well as the resampling of the training data. The main averaged results for the individual tasks are summarized in Figures 5, 6 and 7, and discussed in the following paragraphs (for individual test curves with uncertainties, please refer to Figures S2 S3, S4 and S5 in the supplementary material that accompanies this paper). We choose to show the results from models with an extra CNN layer in the application layers here, because they perform better. Results from the models without the CNN layer in the application layers are in Figure S6 through Figure S11.

From Figures 5, 6 and 7, we can see some interesting common trends. (1) As expected, with more training data the performance of the trained models is better. (2) In almost all test cases, training the encoder plus application layers performs poorer than freshly training a model with the same structure directly using the available training dataset. The only exception is when training data is small for the two classification problems, especially when the training dataset is limited to less than 1500 waveforms. This indicates that the features extracted from the autoencoders, though generic to the waveform itself, may not optimally extract all relevant information for specific applications. When the training dataset is small, the features extracted from the encoders provide additional information to those extract directly from the apparently undertrained baseline model, thus we see better performance. (3) Generally, overcomplete models outperform the undercomplete models (the green and orange solid lines work better than the corresponding dotted lines) in all panels of (c) and (d). Since the bottleneck extracted feature maps are the building blocks for various applications, the models that extract more of them (the overcomplete models) are more likely to include features that are useful to subsequential applications. (4) Overall, the training approach that fine tunes all the layers outperform the approach that only updates the last application dense layers, i.e., the better performance of the green lines if compared with the orange lines. This makes sense, because fine tuning all the encoder layers helps the feature extraction layers adapt to new data sets and different applications. (5) It is not clear whether the bottleneck dimension 34 is better than the 68. Though the solid lines in all panels (a) and (b) are slightly better than the dotted lines, the gaps are small. The next few paragraphs will go over these figures individually, and highlight their differences.

Figure 5 shows the test results for the noise vs. earthquake classification. First, from panels (a) and (b), we can see that using bottlenecks of 34 or 68 has larger effect on the undercomplete models (the gaps between the same color solid and dotted lines). Also, when the training data is increased, the undercomplete models improve little compared to overcomplete models. We think this is due to the smaller number of learned feature maps in the undercomplete models, which limits their performance. For panels (c) and (d), we can see that for the overcomplete models, even when only the application layers are tuned, the performance can be better than if all the encoder layers are tuned in the undercomplete models. This is additional evidence that the overcomplete models can learn more features than the undercomplete models. We also can see from these panels that the performance improvement initially grows faster, but enters into a slow growing area and then into a plateau when the training data size is above 15,000.

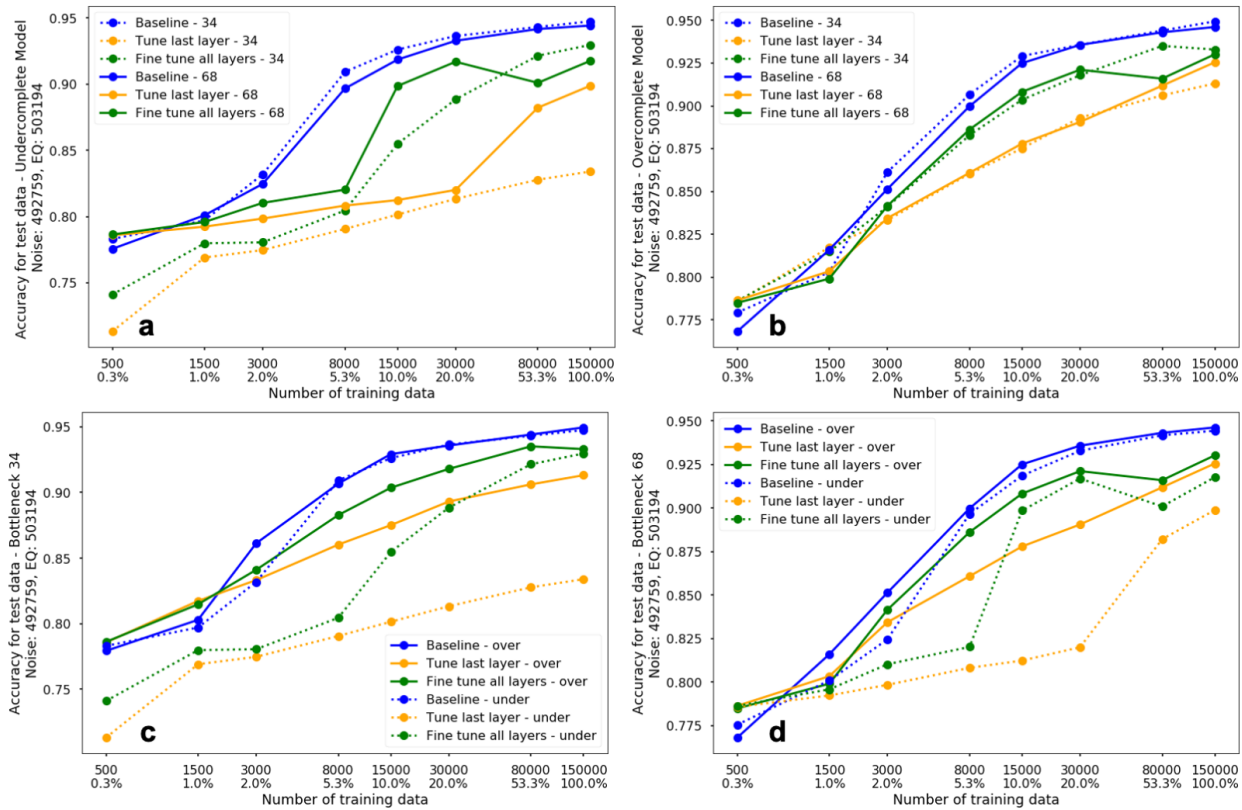

**Figure 5.** Averaged accuracy for the test data set for binary classification, i.e. noise vs. earthquake, with designed models trained against increasing training data (the corresponding training data percentages are also shown with the maximum number of data used as 100%). The test data size for noise samples is 492,759, while the earthquake data size is 503,194. Each data point represents

the average of five training runs with different sampling of training data and new initiation of all weights. Panel (a) compares the results for the undercomplete models, while (b) shows the results for the overcomplete models. Different colors represent different training method, solid and dotted lines are for bottleneck dimensions 68 and 34. Panels (c) and (d) show the results from the bottleneck dimension point of view, with (c) comparing models with bottleneck feature map dimension 34 and (d) with dimension 68. Different colors represent different training method, solid and dotted lines are for overcomplete and undercomplete models, respectively. The x axis is in log scale.

Similarly, Figure 6 summarizes the explosion vs. earthquake classification accuracy for the test dataset. Though it is a similar classification problem as noise vs. earthquake, the essential features that distinguish the two classes are substantially different, with more subtle features between explosion and earthquake waveforms. In panel (b), the performance of the overcomplete models are very similar. This indicates that the encoders from the overcomplete models did a good job of extracting the features that can be used to distinguish the earthquake and explosions, and thus fine tuning all the layers didn't improve the results. In this application, we also do not see the performance plateau as before and we observe that the accuracy improvement is almost linear in the log scale.

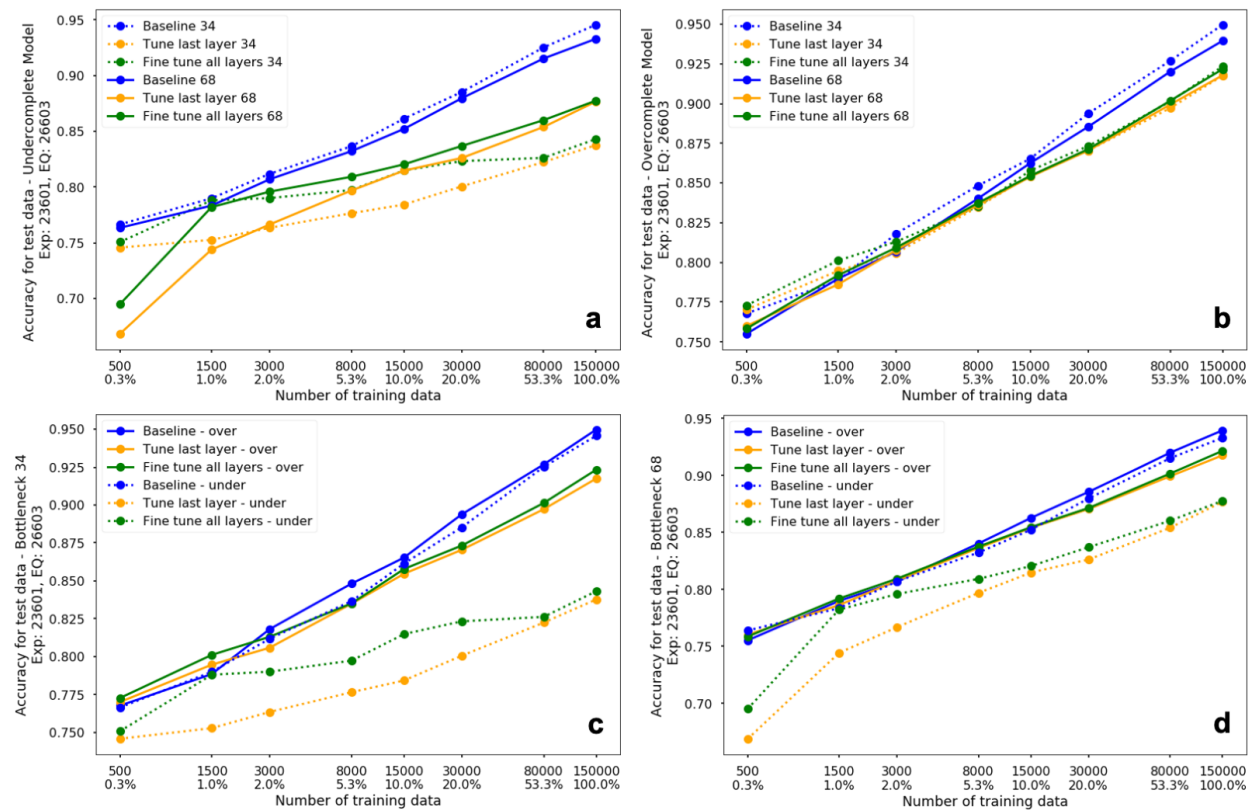

**Figure 6.** Averaged accuracy results on the test dataset for binary classification, i.e. explosion vs. earthquake, with the autoencoder based models trained against different amount of training data (the percentages of the training data are also shown with the maximum number of data used as 100%). The test data sizes for explosion and earthquake are 23,601 and 26,603 respectively. Panels (a) compares the results for the undercomplete models, while (b) shows the results for the overcomplete models. Different colors represent different training methods, solid and dotted lines

are for bottleneck dimensions 68 and 34. Panels (c) and (d) show the results from the bottleneck dimension point of view, with (c) comparing models with bottleneck feature map dimension 34 and (d) with dimension 68. Different colors represent different training method, solid and dotted lines are for overcomplete and undercomplete models. The x axis is in log scale.

Figure 7 shows the test results for the regression problem, i.e., the estimate of the P wave arrival. The features used in this problem are more localized than the previous two examples. We used the standard deviation of the errors as a measure of performance. With sufficient data, the mean of the error distribution approaches zero (see panel d in Figures S2 to S5 in the supplementary material), thus the standard deviation is a good approximation of the performance. We can see from Figure 7 (d), the performances of the designed encoder plus application layers in the overcomplete models with feature map dimension 68 do not exceed the performance of the baseline models as the previous two examples of classification when training dataset is small, but the difference is not large. In Figure S3(c), we can also see that the shaded area (the standard deviation of the 5 tests) for the green line has regions lower than the blue baseline, which means that there are cases among the five runs that performed better than the baseline model. We can also see that with more training data available, the performance of the models with a fine tuning of all the layers are getting closer to the baseline performance, until there is a constant gap. One interesting thing in panel (c) and (d) of Figure 7 is that, when data sizes are large, the undercomplete models with a fine tuning of all the layers have a comparable performance to that of the overcomplete models, unlike the two classification cases. We attribute this to relevant features being more localized in the P phase picking problem and can be captured with fewer feature maps. Therefore, more feature maps in the overcomplete models do not necessarily improve the results if compared with the undercomplete models with a fine tuning of all layers.

From Figure 7, we can see the errors still seem to be decreasing. Since we have more training data in the STEAD, we continued the training with more data up to 300,000. Figure 8 shows the performance of the models on the test data with more training data (note, in this case, the x-axis is linear scale to avoid label overlap). As expected, the green line (fine tune of all layers) and blue line (baseline) in Figure 8 shows further improvement, although the improvement rate is smaller compared with training data size of 100,000. Besides, the gap between the green and blue line continues to decrease. Figure 9 also shows the distributions of estimated errors (predicted time – labeled time) with different training data sizes. We can see that with more training data available, the performance of the models with the fine tuning of all the layers are approaching to the baseline model. We also see that the performance of the model with a fine tuning of only the last layer improves slowly.

We also tested to see if the SNR (signal to noise ratio) will affect the results, i.e., whether the autoencoder-based model can outperform the baseline model when SNR is low (or noisier data) using a large amount of training data. Figure S26 shows the STD of the absolute errors versus SNR on the test data, and we do not see the autoencoder based model perform better for noisy data. As shown in figure 7, when training data is small, the autoencoder based model will perform slightly better than the baseline model trained from scratch. This can be seen in figure S27 as well.

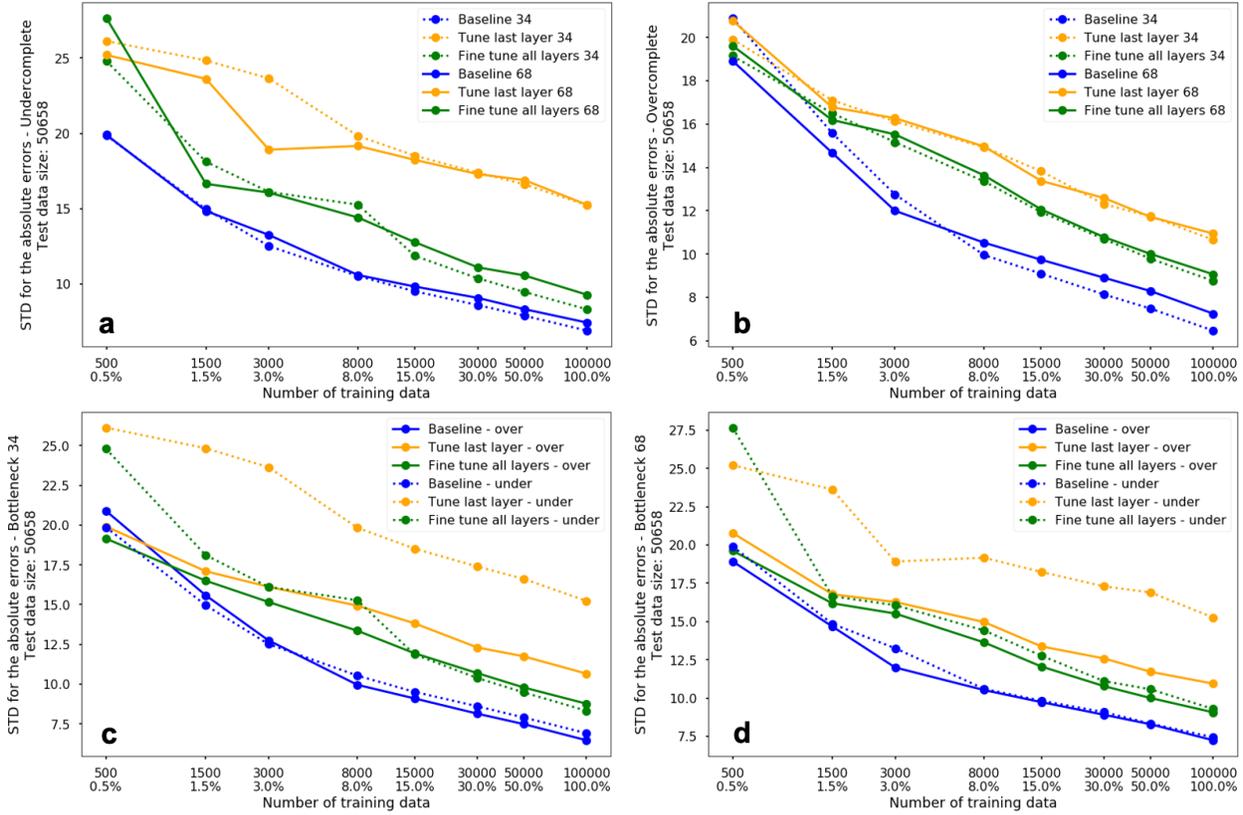

**Figure 7.** Averaged standard deviation of the absolute estimation errors for the regression problem, i.e. P arrival estimation, with autoencoder based models trained against different amounts of training data (the percentages of the training data are also shown with the maximum number of data used as 100%). Test data size is 50,658. Panels (a) compares the results from the undercomplete models, while (b) shows the results from the overcomplete models. Different colors represent training methods, solid and dotted lines are for bottleneck dimensions 68 and 34. Panels (c) and (d) show the results from the bottleneck dimension point of view, with (c) comparing models with bottleneck feature map dimension 34 and (d) with dimension 68. Different colors represent training method, solid and dotted lines are for overcomplete and undercomplete models. The x axis is in log scale.

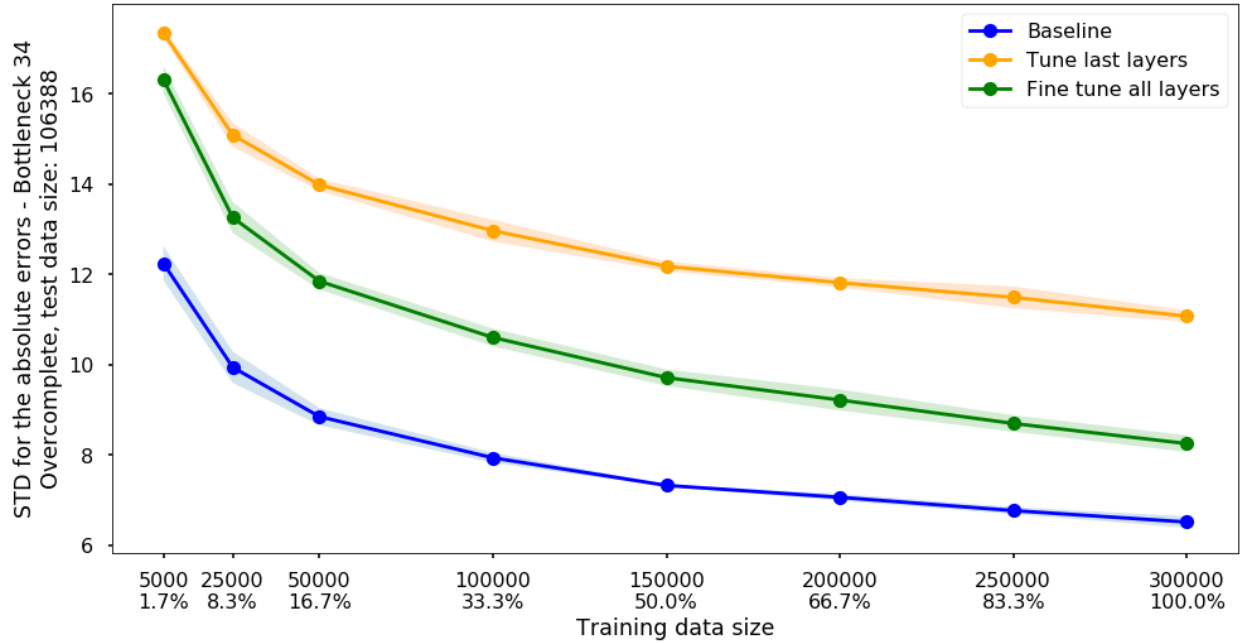

**Figure 8.** Standard deviation of the absolute error when training with more data for the P wave arrival estimation using STEAD (Magnitude >= 1.5 and distance within 200 km). Each dot is the mean value of the five models trained with different initialization of weights and randomly sampled training data, the shaded areas represent one standard deviation. The x axis is in linear scale.

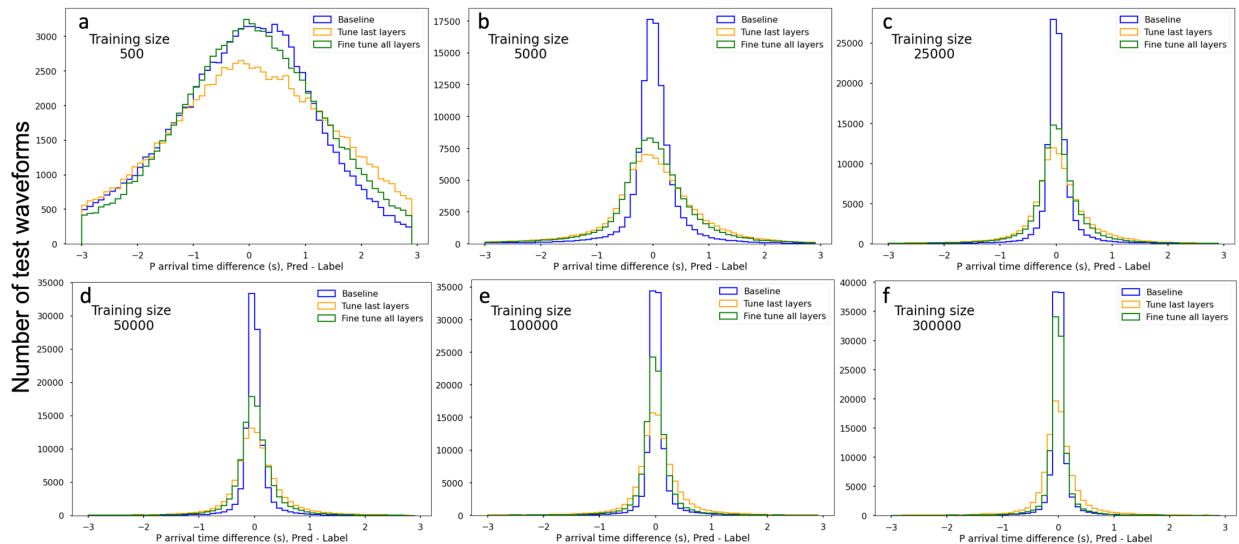

**Figure 9.** P wave arrival time error (Prediction – Label) distribution on test data with different training data size, the test data size for each panel is 106,388. Test error distribution with (a) training data size 500; (b) training data size 5,000; (c) training data size 25,000; (d) training data size 50,000; (e) training data size 100,000; (f) training data size 300,000.

## 4.3 Computational cost

Figure 10 shows the computational cost of training different types of models with various amounts of training data. Due to training process takes most of the time, while the test only takes fractions of seconds to run on a single waveform, we only show the computation cost of the training here. Times were measured on two Nvidia Quadro RTX 6000 GPUs from the beginning of the training until the model convergence (the validation accuracy or loss did not improve for 20 epochs). The timing roughly increases exponentially. When the data sizes are relatively small, the different methods have similar timing cost (or small differences). In fact, many of the encoder plus the application layer models converge faster than the baseline model (see Figure S16 in the supplementary material for more details). When data sizes are becoming larger, roughly around 8,000, we start to see that the training times increase dramatically, especially for the encoder schema with fine tuning. Overall, as expected, a fine tuning of all the layers with small learning rate takes the longest time to converge, while training the baseline model consumes the least time in all these models.

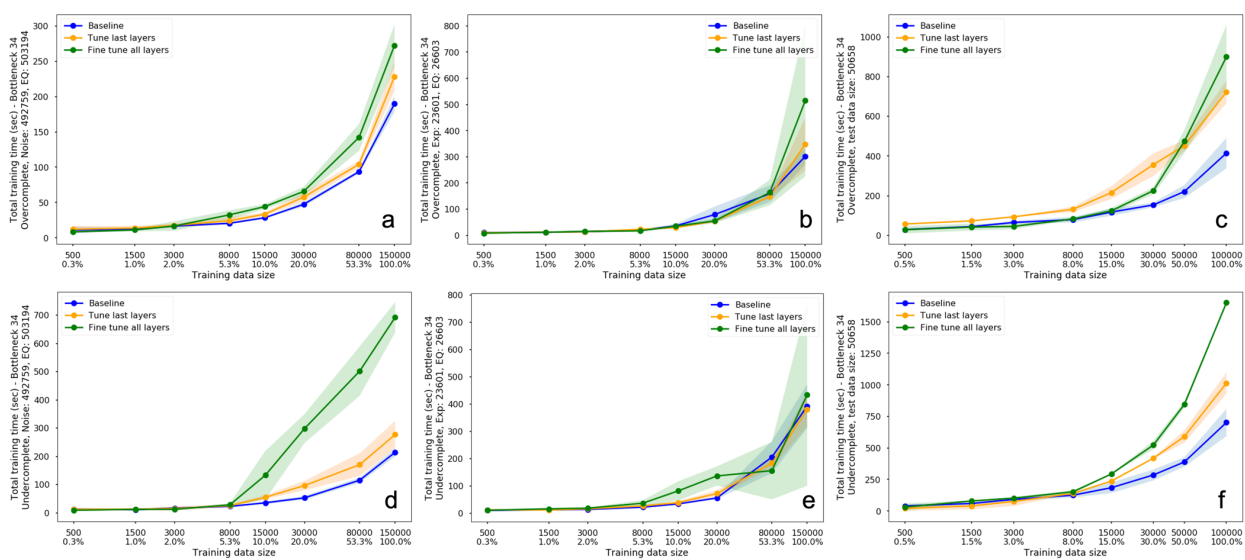

**Figure 10.** Total time in seconds for the models to converge (if the validation performance does not improve for 20 epochs, the training process stops), dots are mean values and shaded areas are one standard deviation from the five runs. The models were trained on 2 Nvidia Quadro RTX 6000 GPUs. Overcomplete models are shown in the top row panels (a), (b) and (c), while undercomplete models are shown in the bottom row panels. Noise vs. earthquake problem is shown in the first column, panels (a) and (d). Explosion vs. earthquake problem is shown in the 2nd column, panels (b) and (e). P arrival estimation problem is shown in the last column, panels (c) and (f). We used a fixed batch size of 256 in all the tests for comparison purposes.

## 5 Conclusions

We evaluate the idea of using an autoencoder as a generic feature extractor for different seismological applications. The main conclusion is the autoencoder using the overcomplete model structure with an extra convolutional layer in the application layer is achieving the best performance when trained with the fine tuning of all layers. In most cases the autoencoder approach does not outperform a baseline model that is tailored to the specific application.

However, the autoencoder-based models can perform slightly better than a baseline model when the training dataset is small.

## 6 Acknowledgments, Samples, and Data

The views expressed in the article do not necessarily represent the views of the U.S. Department of Energy or the U.S. Government. This research was funded by the National Nuclear Security Administration, Defense Nuclear Nonproliferation Research and Development (NNSA DNN R&D). The authors acknowledge important interdisciplinary collaboration with scientists and engineers from Los Alamos National Laboratory (LANL), Lawrence Livermore National Laboratory (LLNL), Mission Support and Test Services, LLC (MSTS), Pacific Northwest National Laboratory (PNNL), and Sandia National Laboratories (SNL). This research was performed in part under the auspices of the U.S. Department of Energy by the LLNL under Contract Number DE-AC52-07NA27344. This is LLNL Contribution LLNL-JRNL-828227. We also thank the authors of the datasets used in this study, including STEAD, LEN_DB, SPE, BASE, and iMush, especially we thank Brandon Schmandt and Ruijia Wang for providing a list of stations for downloading data for SPE (https://www.nnss.gov/docs/fact_sheets/NNSS-SPE-U-0034-Rev01.pdf), BASE (https://www.passcal.nmt.edu/content/bighorn-arch-seismic-experiment-results), and iMush (http://geoprisms.org/education/report-from-the-field/imush-spring2015/). We also thank IRIS (https://www.iris.edu/hq/) to host the seismic data for research purposes. We thank useful discussions from Bill Walter at the Lawrence Livermore National Laboratory. All the analysis are done in Python and the deep learning framework used here is TensorFlow (Abadi et al., 2016), and seismological related analysis are using Obspy (Beyreuther et al., 2010; Krischer et al., 2015), we thank the awesome Python communities to make everything openly available.

# Supplementary Materials: Evaluation of Using Deep Convolutional Autoencoder as Generic Feature Extractions in Seismological Applications

**Qingkai Kong[1], Andrea Chiang[1], Ana Cristina[1] Aguiar Moya[1], M. Giselle Fernández-Godino [1], Stephen C. Myers[1], Donald D. Lucas[1]**

[1] Lawrence Livermore National Laboratory.

Corresponding author: Qingkai Kong (kong11@llnl.gov)

The views expressed in the article do not necessarily represent the views of the U.S. Department of Energy or the U.S. Government. This research was funded by the National Nuclear Security Administration, Defense Nuclear Nonproliferation Research and Development (NNSA DNN R&D). The authors acknowledge important interdisciplinary collaboration with scientists and engineers from Los Alamos National Laboratory (LANL), Lawrence Livermore National Laboratory (LLNL), Mission Support and Test Services, LLC (MSTS), Pacific Northwest National Laboratory (PNNL), and Sandia National Laboratories (SNL). This research was performed in part under the auspices of the U.S. Department of Energy by the LLNL under Contract Number DE-AC52-07NA27344. This is LLNL Contribution LLNL-JRNL-828227.

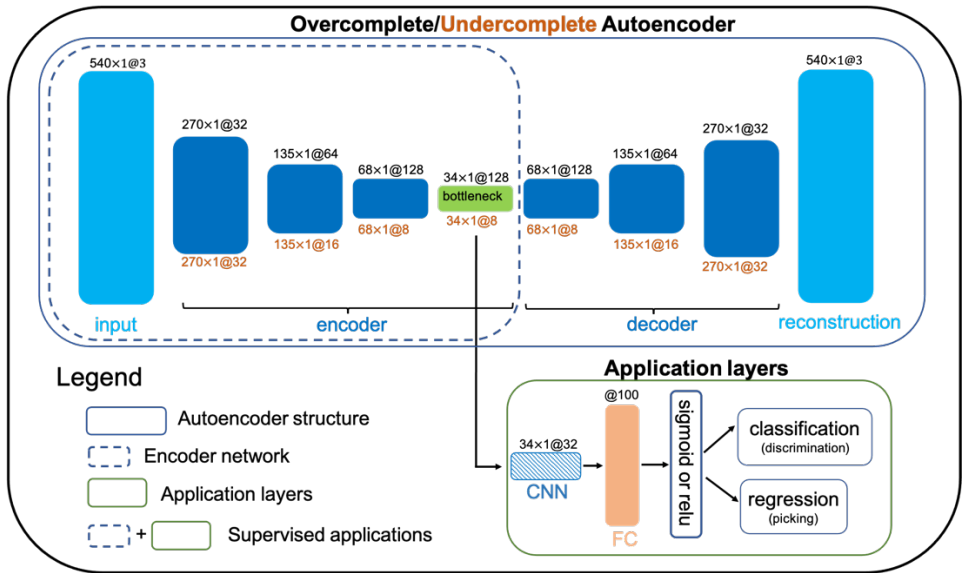

**Figure S1** The workflow of the experiments but with the bottleneck layer dimension is 34 comparing with the 68 in the main text. For the one with the bottleneck layer dimension is 17, we just add another layer to shrink the size to 17×1@8 (undercomplete) or 17×1@128 (overcomplete) followed by a scaling up 34×1@8 (undercomplete) or 17×1@128 (overcomplete).

**Test results for the designed models with the CNN layer in the application layers**

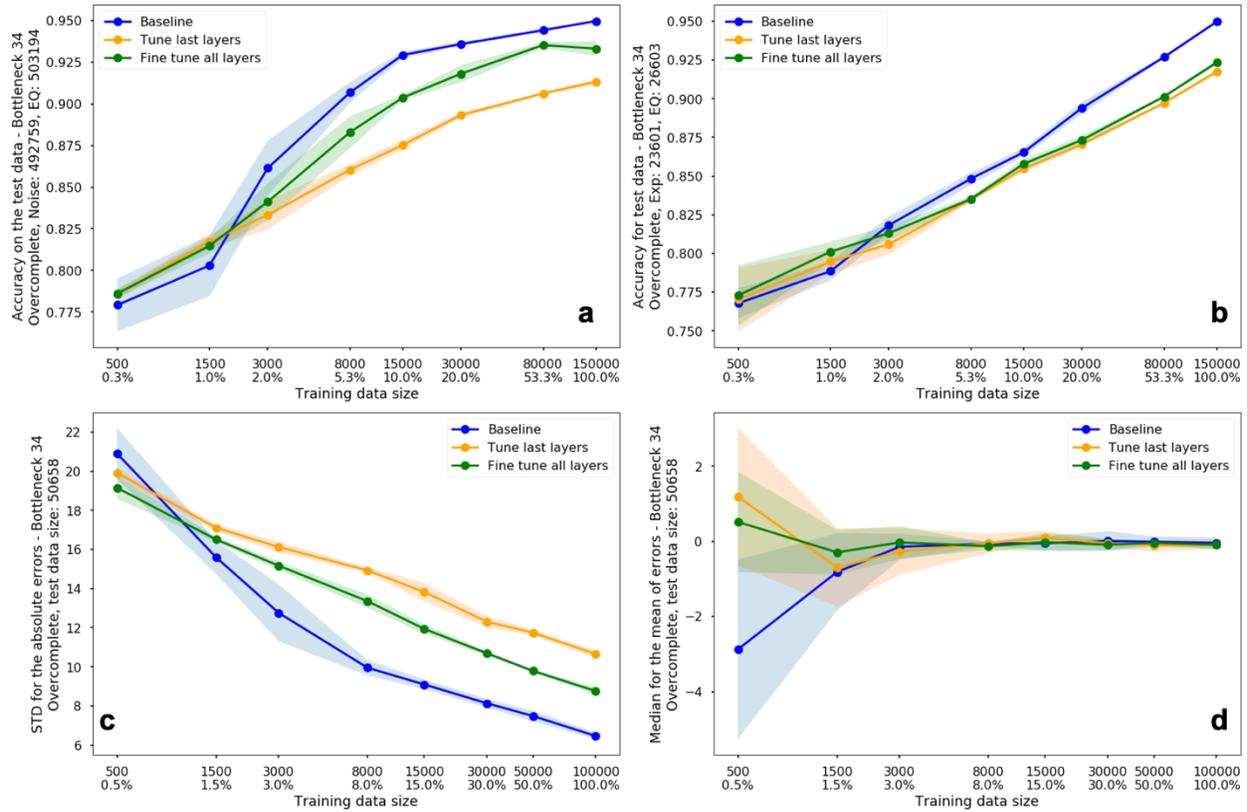

Figure S2 Test performance for the overcomplete model with bottleneck 34 dimensions. The dots are average results, with shaded areas are the standard deviation. (a) accuracy for the noise vs. earthquake classification, (b) accuracy for the explosion vs. earthquake classification, (c) standard deviation of the absolute errors for the P wave arrival estimation, (d) Median for the mean of the errors for the P wave arrival estimation.

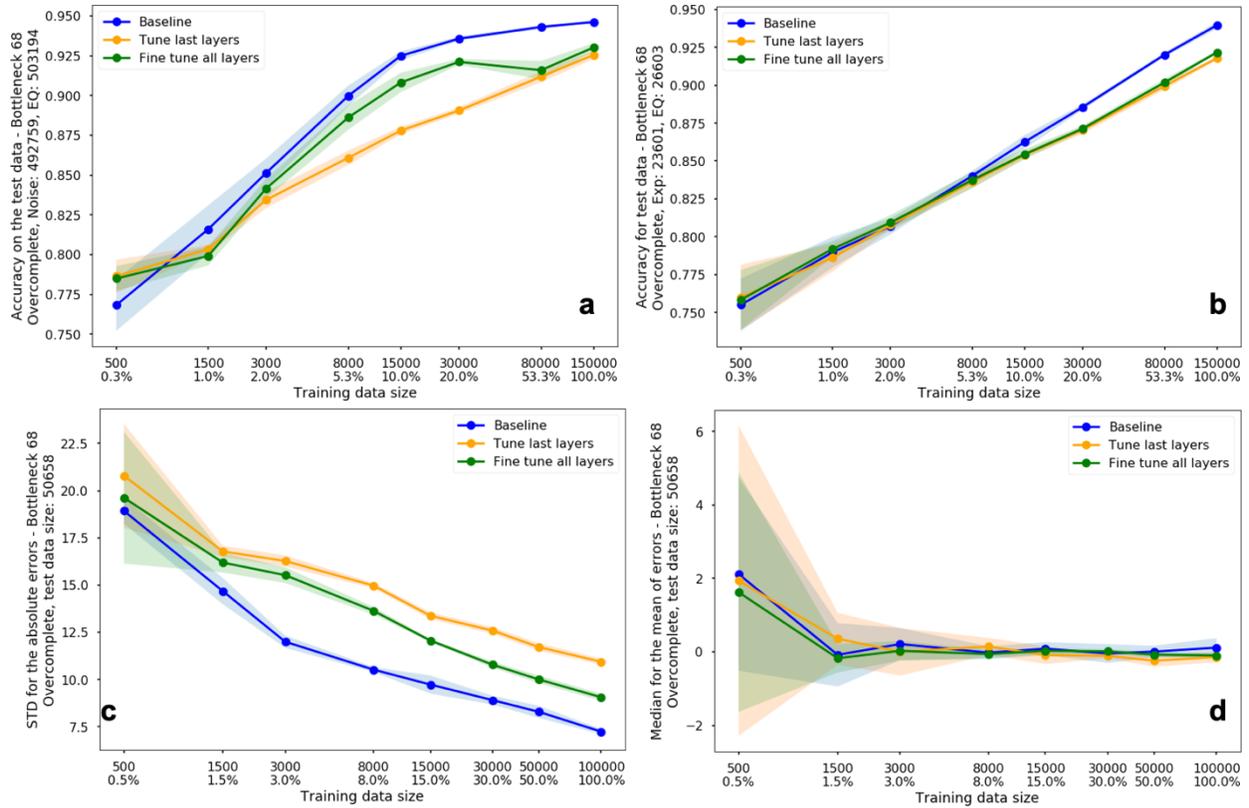

Figure S3 Test performance for the overcomplete model with bottleneck 68 dimensions. The dots are average results, with shaded areas are the standard deviation. (a) accuracy for the noise vs. earthquake classification, (b) accuracy for the explosion vs. earthquake classification, (c) standard deviation of the absolute errors for the P wave arrival estimation, (d) Median for the mean of the errors for the P wave arrival estimation.

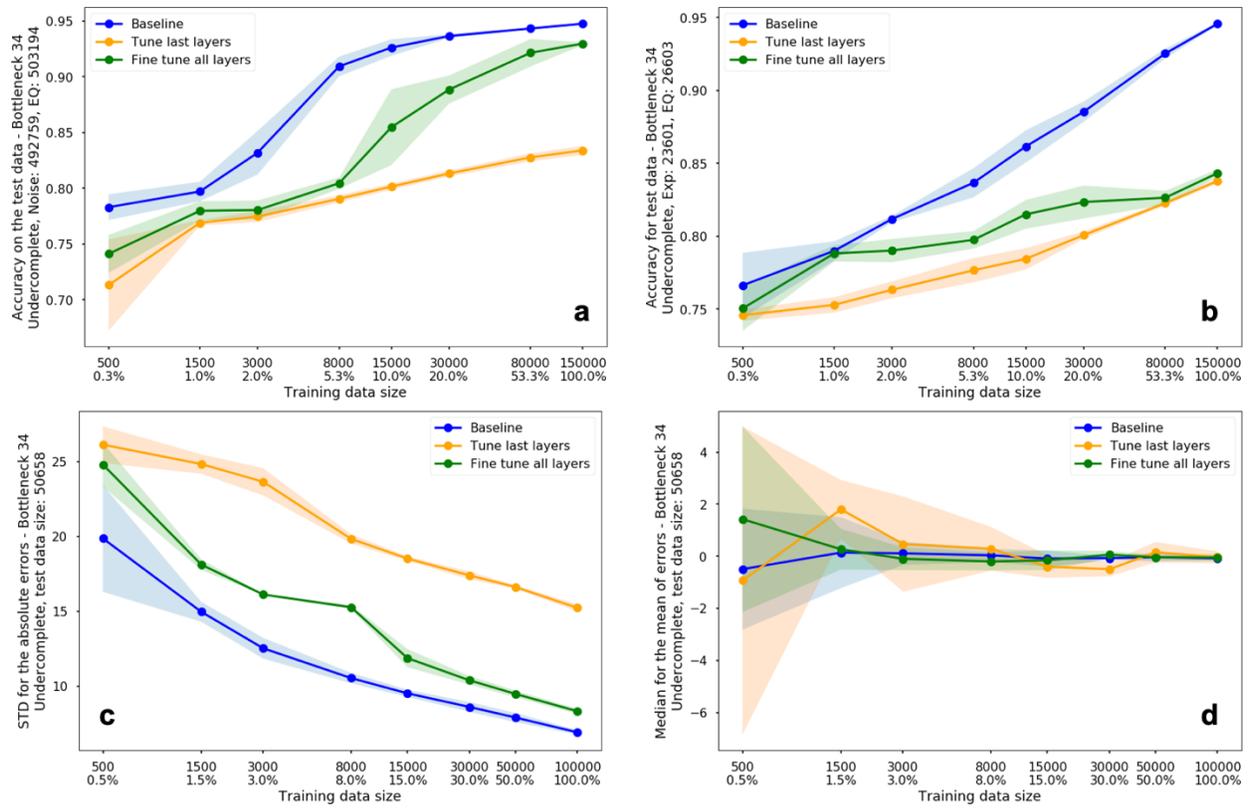

Figure S4 Test performance for the undercomplete model with bottleneck 34 dimensions. The dots are average results, with shaded areas are the standard deviation. (a) accuracy for the noise vs. earthquake classification, (b) accuracy for the explosion vs. earthquake classification, (c) standard deviation of the absolute errors for the P wave arrival estimation, (d) Median for the mean of the errors for the P wave arrival estimation.

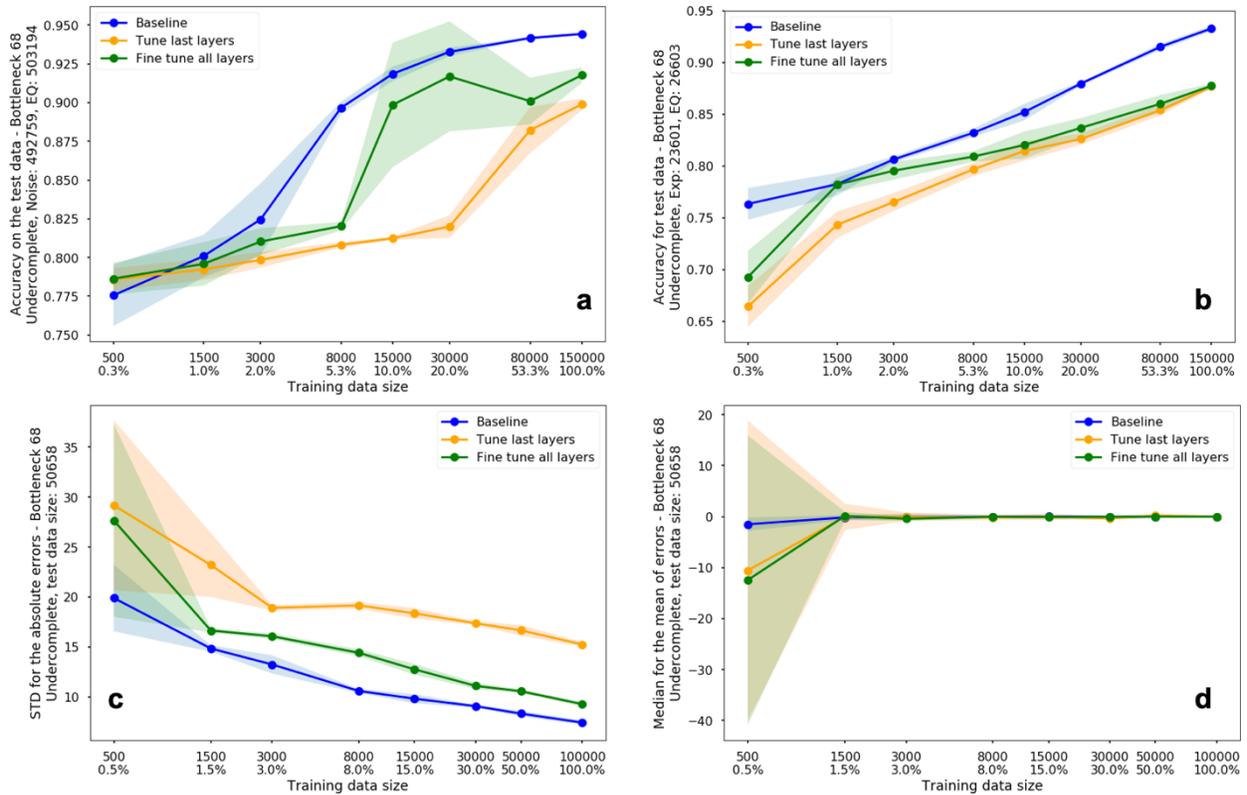

Figure S5 Test performance for the undercomplete model with bottleneck 34 dimensions. The dots are average results, with shaded areas are the standard deviation. (a) accuracy for the noise vs. earthquake classification, (b) accuracy for the explosion vs. earthquake classification, (c) standard deviation of the absolute errors for the P wave arrival estimation, (d) Median for the mean of the errors for the P wave arrival estimation.

**Test results for the designed models without the CNN layer in the application layers**

We tested the models when we have the bottleneck feature map dimensions as 68 and 34 with different amount of training data. For each model specific configuration, we ran 5 different training/testing with varying the initialization of the model weights as well as the resampling of the training data. The main averaged results for the individual tasks are summarized in figure S6 – S8, and discussed in the following paragraphs. From these 3 figures, we can see some interesting common trends: (1) As expected, with more training data, we can expect the performance of the trained models are getting better. (2) The baseline model that training from scratch almost outperform all the other models combining the encoder layers with the application layer, except for a few cases when training data size is small. This indicates that the features extracted from the autoencoders, though generic to the waveform itself, may not very well target specific applications. (3) For the overcomplete model, either the 34 or 68 bottleneck dimensions model, the performance of only training the last layer or fine tune all layers are quite similar. One possible explanation is the features extracted by these models have enough explain power for these different applications to achieve good (not perfect) solutions. Even though the 68-dimension autoencoder has lower reconstruction error than the 34-dimension model as shown in figure 2, the more feature dimensions it has doesn't necessarily provide new information for these applications. Put it another way, the 34-dimension model, though with smaller features extracted, has almost the same explanation power due to the large amount of feature maps we were using. (4) Overall, the approach 2 that fine tune all the layers have better test results than that from approach 1, only tune the last application layer. This makes sense, because fine tune all the encoder layers help the feature extraction layers more adapt the new cases in different applications. In the next few paragraphs, we will go over the individual details in these figures and highlight the differences.

Figure S6 shows the test results for the task of noise vs. earthquake classification. First of all, we can see from panel (a) that the shallower models perform better (solid lines higher than dotted lines) for the undercomplete model. When having very few data, such as 500 training samples, the performances of approach 1 and 2 are both close or slightly higher than that from the baseline model when we tuned all the layers (green lines). With more training data present, the performance of the baseline model increases fast, while that from tuning only the last layer has only small increasement, but the model which has the fine tune of all the layers increases slowly first, then catches up the baseline when data size is sufficient. The relative flat orange lines in the undercomplete model shows that increasing the training data doesn't improve the model performance too much, we think the reason is due to the small number of bottleneck features that extracted in the undercomplete models. When comparing the performance of undercomplete and overcomplete models in panel (c) and (d), we can see that the overcomplete models generally perform better than the undercomplete models, though when training data size is very small or large, the undercomplete models have similar or slightly better results.

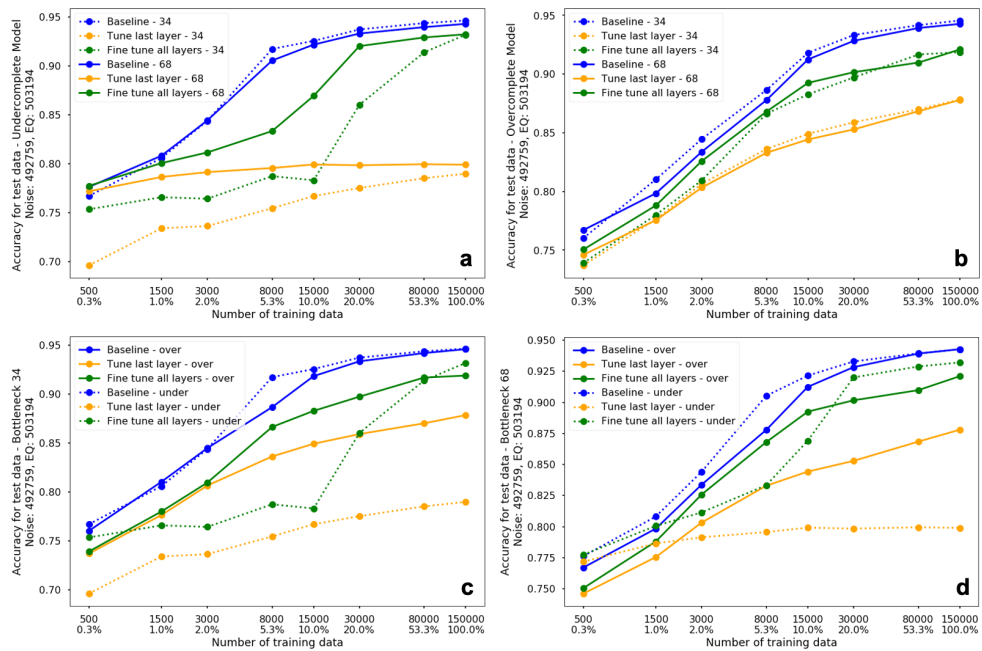

**Figure S6**. Averaged test results for noise vs. earthquake classification with designed models trained against different amount of training data (the percentages of the training data are also shown with the maximum number of data used as 100%). Each data point represents the average of 5 training runs with different sampled training data and new initiation of all the weights. (a) and (b) Comparison of models with bottleneck dimensions as 34 and 68 for undercomplete and overcomplete models respectively. (c) and (d) Comparison of overcomplete and undercomplete models with bottleneck dimensions as 34 and 68 respectively. The x axis is in log scale.

Figure S7 shows the test results for the task of explosion vs. earthquake classification. Though it is a similar classification problem as noise vs. earthquake, the essential features that distinguish the two classes are dramatically different, with the features are more subtle between explosion and earthquake waveforms. Thus we see different patterns here in the results. First, the improvement of the accuracy for all the models increases more linearly with the larger training data size (note, the x axis here is log scale, therefore, this linearity is regarding to logarithm size of the data). We also don't see the flatten of the accuracy when used 150,000, which indicates the performance can still improve when adding more data. When training data size is small, the performance differences between the baseline and the two training approaches of the autoencoder based models are very small for the overcomplete model in panel (b). While the opposite results can be seen for the undercomplete model, where the performance gap is larger with small data size. The overcomplete models are all perform better than the underperform models in panel (c) and (d), this is much clear than that in the previous figure.

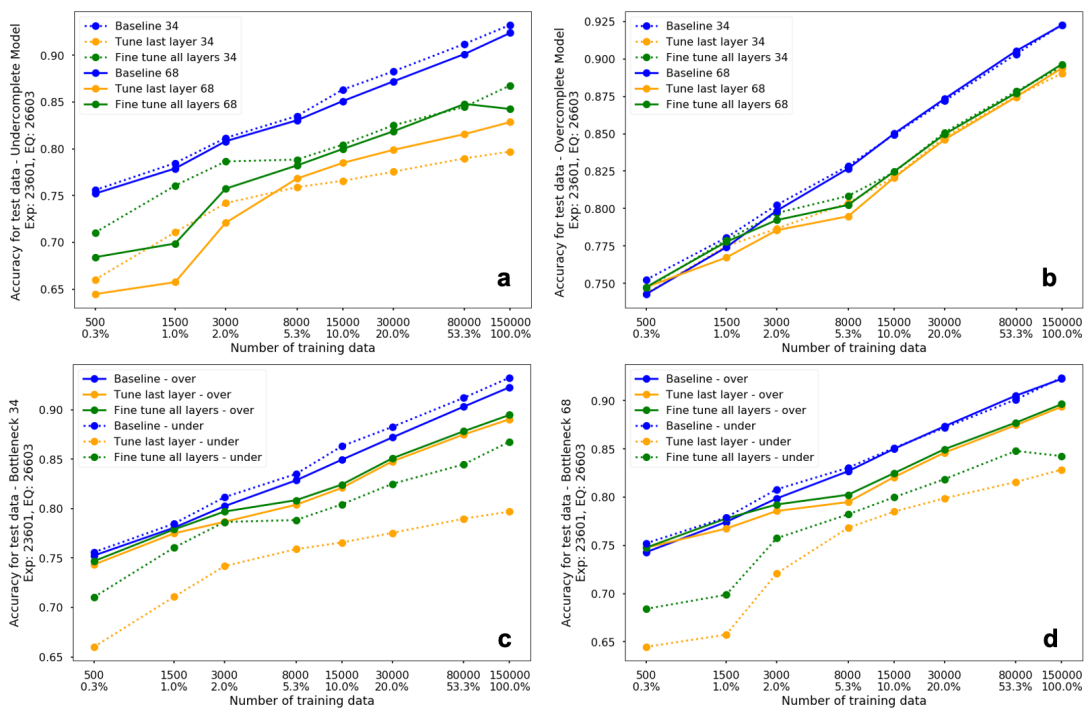

**Figure S7.** Averaged test results for explosion vs. earthquake classification with designed models trained against different amount of training data (the percentages of the training data are also shown with the maximum number of data used as 100%). Please refer to figure 3 for detailed captions for each panel. The x axis is in log scale.

Figures S8 and S9 show the test results for the regression problem, i.e. estimate of the P wave arrival. The features used in this problem are more localized features than the previous two examples. We used the standard deviation of the errors as a proximation of the performance. As with sufficient data, the mean of the error distribution approaching to zero, thus the standard deviation is a good approximation of the performance, with smaller values are better. In figure 5, we can see the undercomplete models have relatively flat orange lines in panel (a), which indicate the encoded features from the reconstruction of the waveforms are not so useful for determine the arrival of the P wave, thus only tuning the last layer doesn't improve the performance much even with large amount of training data. But when we started to tune all the layers, which adjusted the encoded features, performance improving with more training data available. For the overcomplete models in panel (b), when have relatively small or large training data, the green lines have closer performance to the blue lines, especially, we see the gaps between the green and orange lines are increasing when more training data available. This shows that the performance improvement from the adjustment of the extracted features is getting better when more data are provided. Figure S9 also shows the distributions of estimated errors (predicted time – labeled time) with different training data sizes. We can see that with more training data available, the performance of the model with fine tune all the layers is approaching to the baseline model that trained from scratch. But the performance from the model with only tuned the last layer improves slowly.

Figure S10 and S11 shows one example of the performance and timing for training the models with more data up to 300,000 with uncertainties.

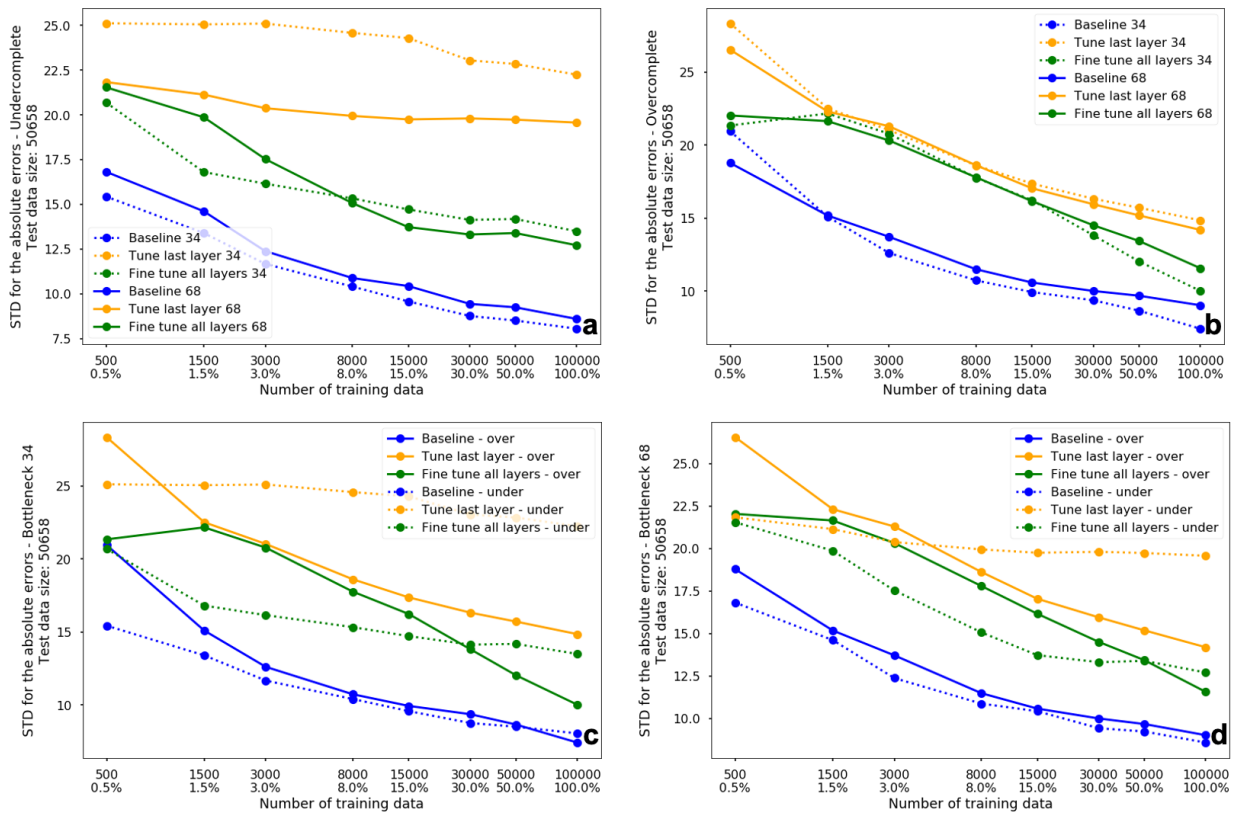

**Figure S8.** Averaged standard deviation of the absolute estimation errors for P arrival estimation with designed models trained against different amount of training data (the percentages of the training data are also shown with the maximum number of data used as 100%). Please refer to figure 3 for detailed captions for each panel. The x axis is in log scale.

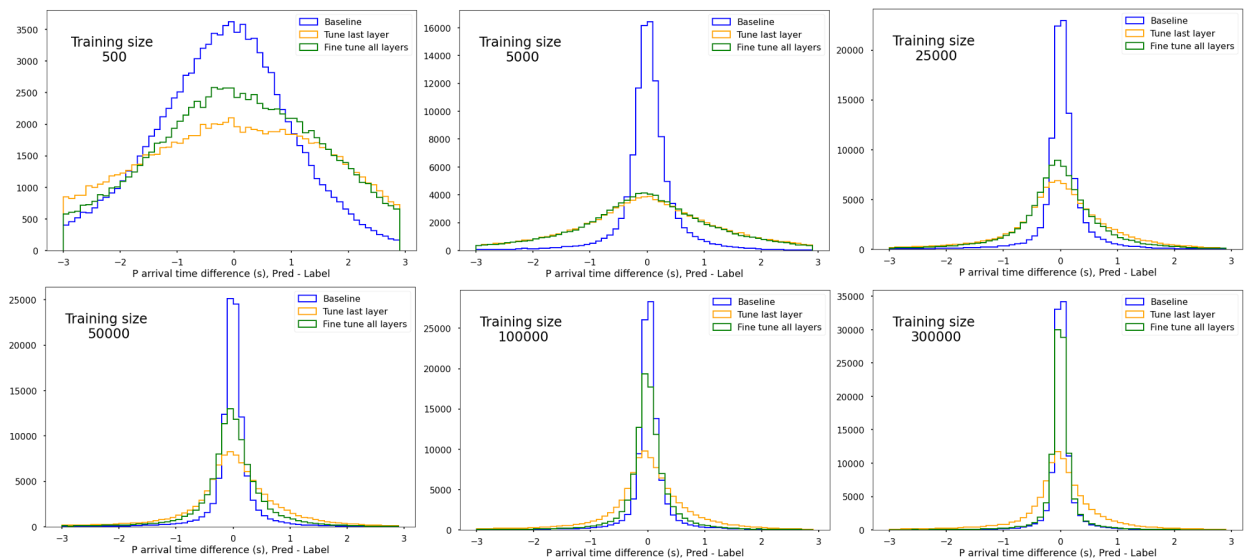

**Figure S9.** P wave arrival time error (Prediction – Label) distribution with different training data size, the test data size for each panel is 106,388.

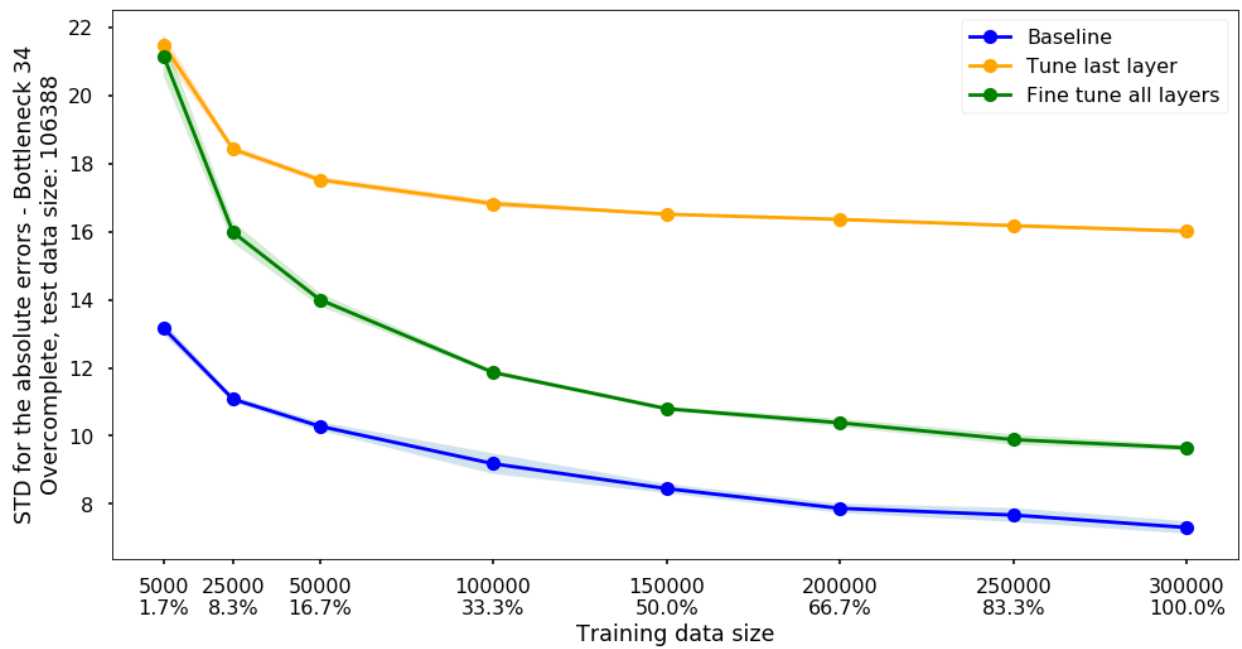

**Figure S10.** Training and testing with more data for the P wave arrival estimation using STEAD (Magnitude >= 1.5 and distance within 200 km). Each dot is the mean value of 5 models trained with different weights initialization and randomly sampled training data, the shaded areas are the standard deviation from the 5 runs. The x axis is in linear scale.

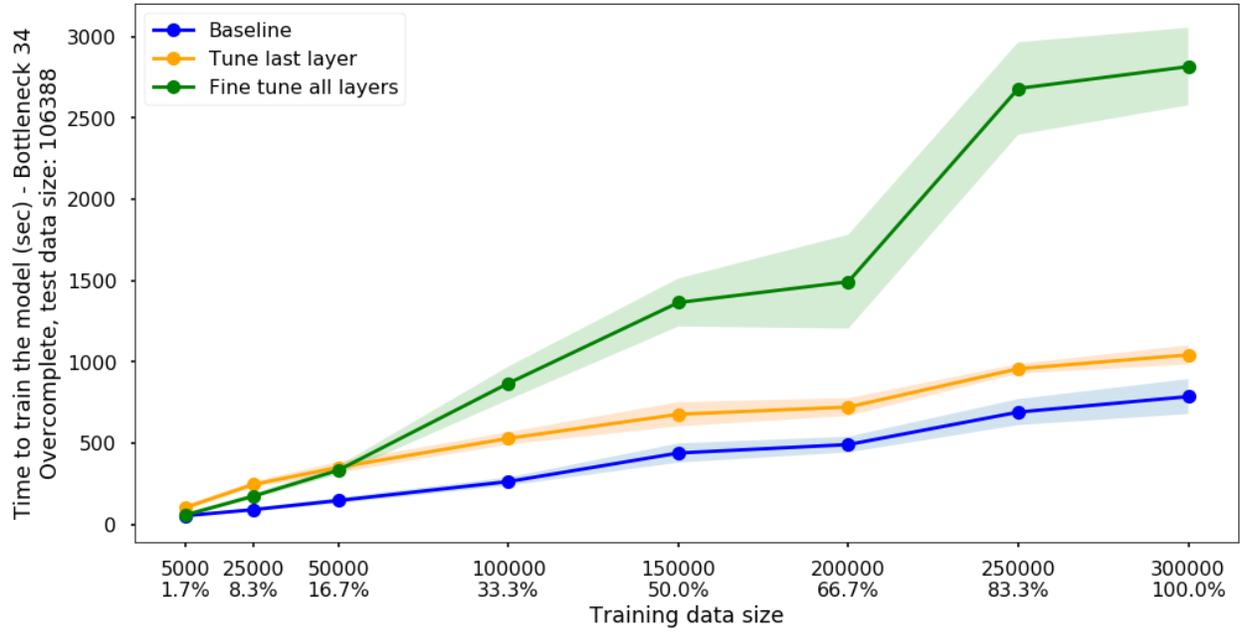

**Figure S11.** The total time in seconds for the model to converge (if the validation performance doesn't improve for 20 epochs, the training process stops), dots are mean values and shaded areas are the standard deviations from the 5 runs. Models were trained on 2 Nvidia Quadro RTX 6000 GPUs.

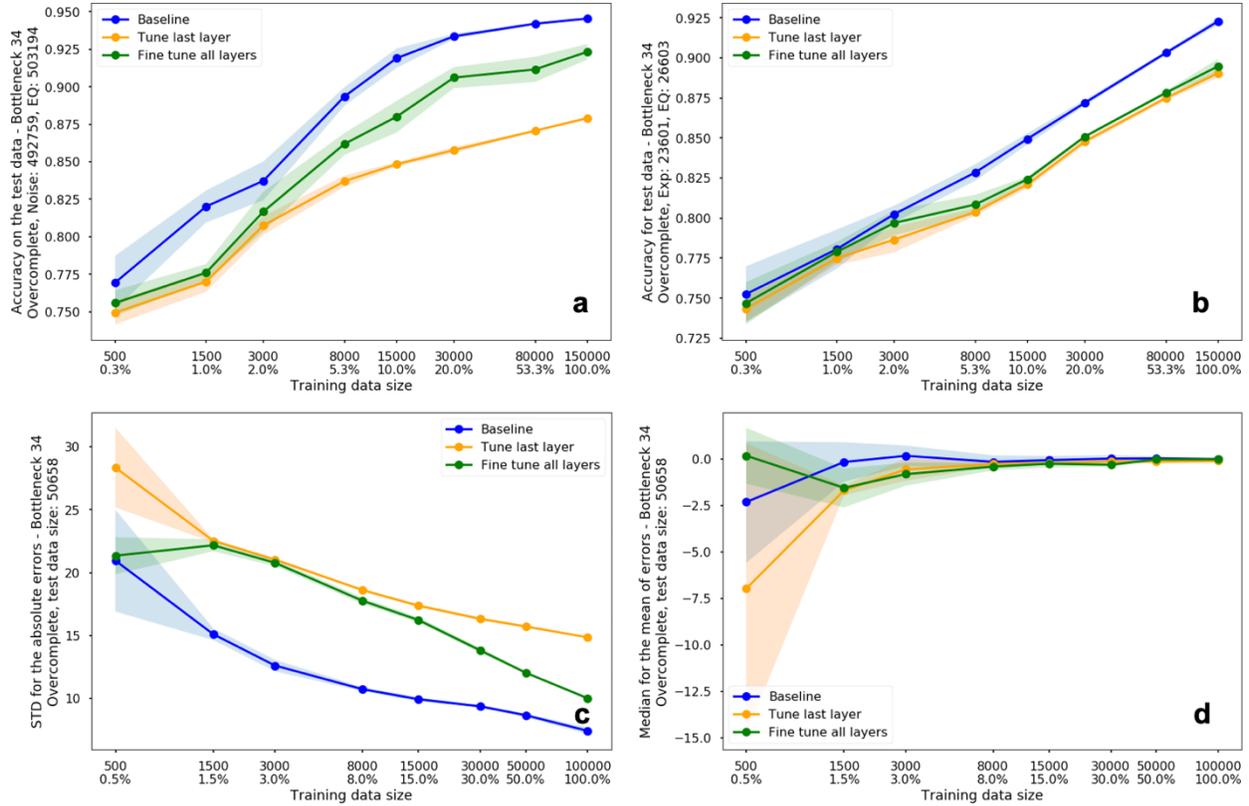

**Figure S12**. Test performance for the overcomplete model with bottleneck 34 dimensions. The dots are average results, with shaded areas are the standard deviation. (a) accuracy for the noise vs. earthquake classification, (b) accuracy for the explosion vs. earthquake classification, (c) standard deviation of the absolute errors for the P wave arrival estimation, (d) Median for the mean of the errors for the P wave arrival estimation.

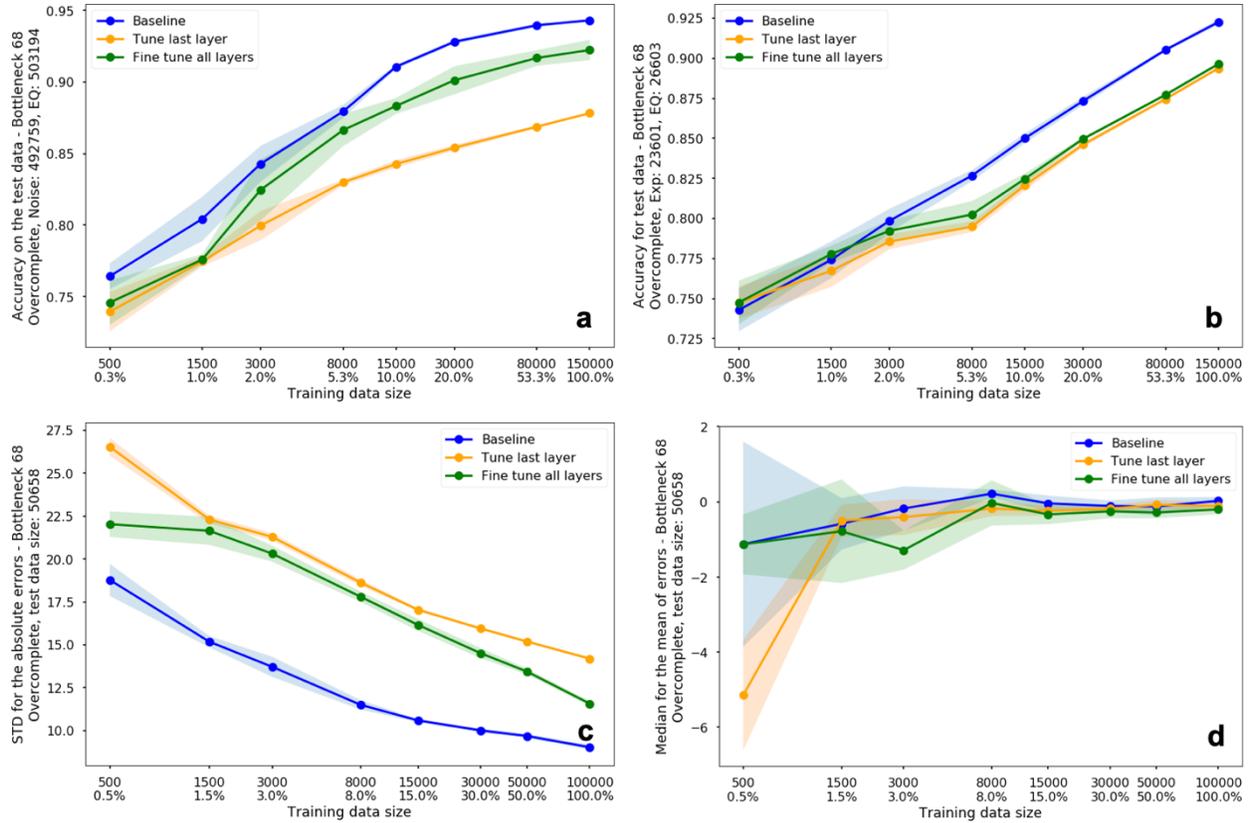

**Figure S13.** Test performance for the overcomplete model with bottleneck 68 dimensions. The dots are average results, with shaded areas are the standard deviation. (a) accuracy for the noise vs. earthquake classification, (b) accuracy for the explosion vs. earthquake classification, (c) standard deviation of the absolute errors for the P wave arrival estimation, (d) Median for the mean of the errors for the P wave arrival estimation.

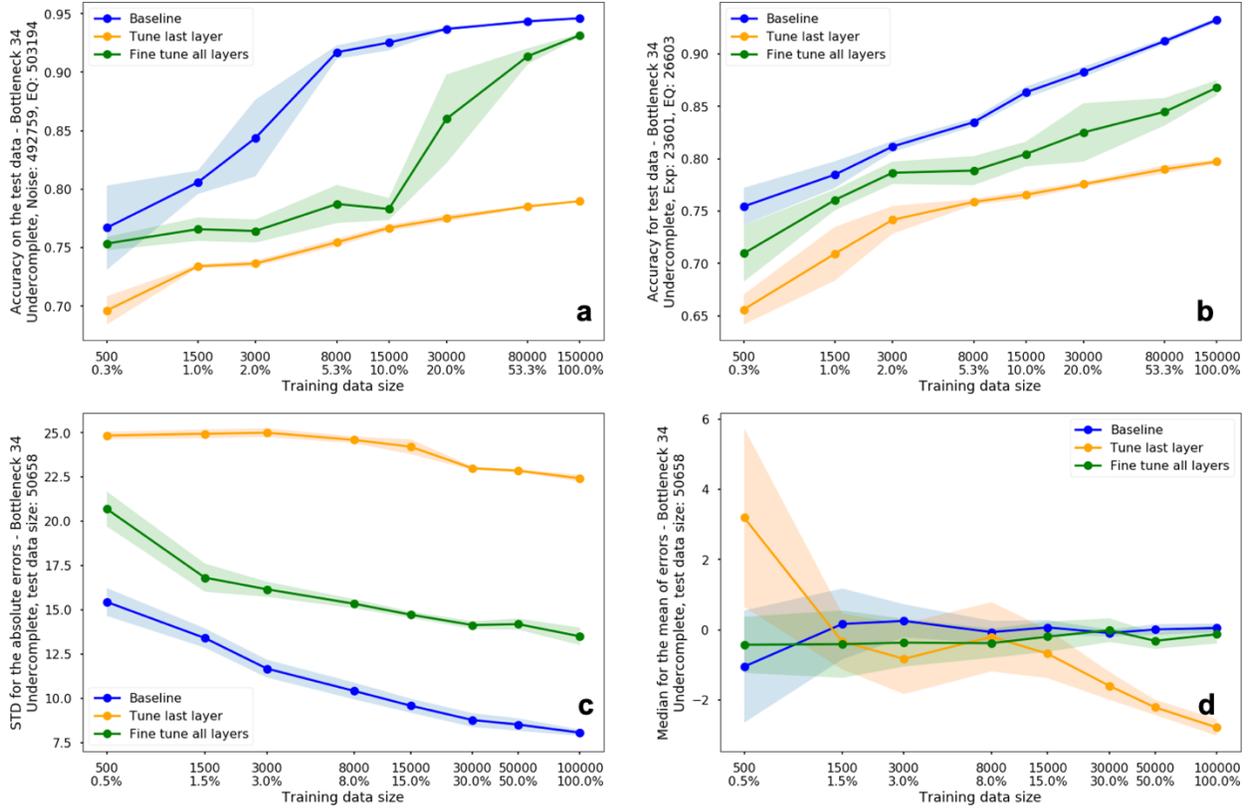

**Figure S14.** Test performance for the undercomplete model with bottleneck 34 dimensions. The dots are average results, with shaded areas are the standard deviation. (a) accuracy for the noise vs. earthquake classification, (b) accuracy for the explosion vs. earthquake classification, (c) standard deviation of the absolute errors for the P wave arrival estimation, (d) Median for the mean of the errors for the P wave arrival estimation.

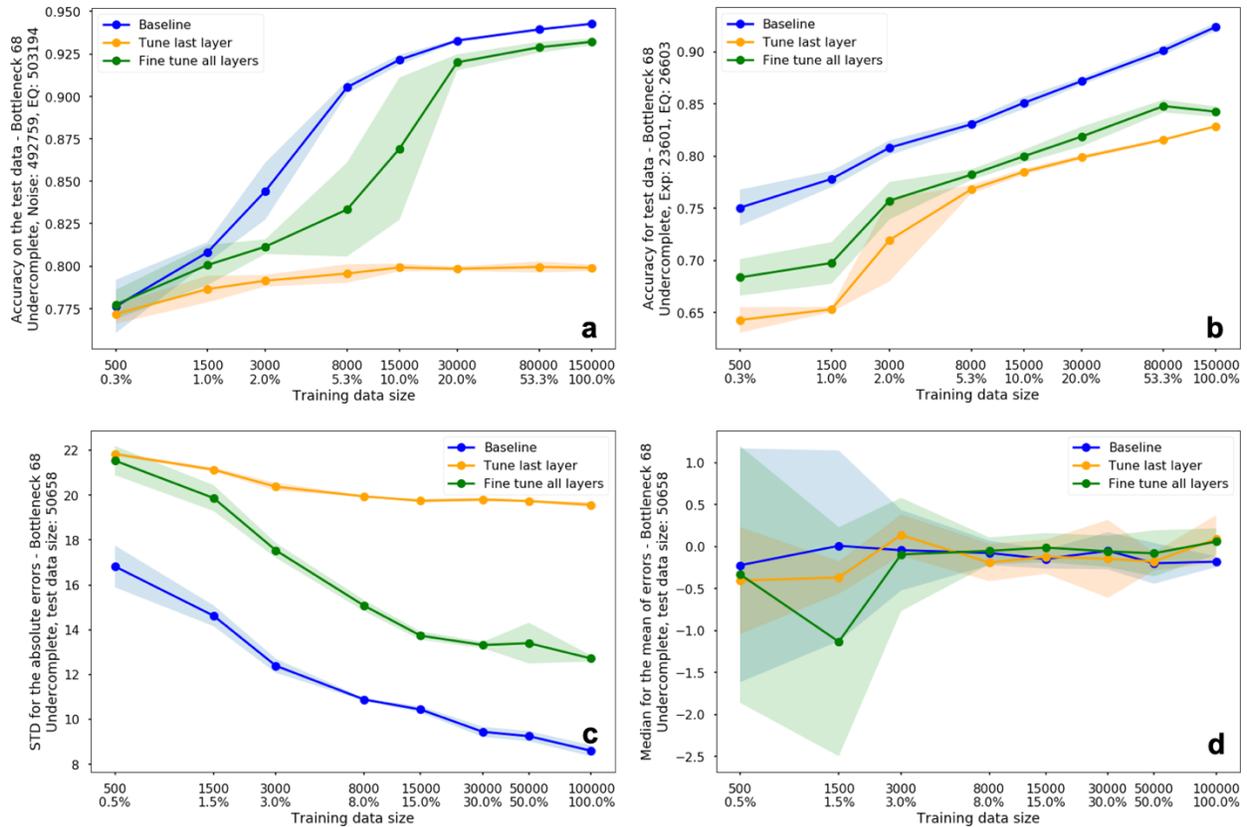

**Figure S15.** Test performance for the undercomplete model with bottleneck 34 dimensions. The dots are average results, with shaded areas are the standard deviation. (a) accuracy for the noise vs. earthquake classification, (b) accuracy for the explosion vs. earthquake classification, (c) standard deviation of the absolute errors for the P wave arrival estimation, (d) Median for the mean of the errors for the P wave arrival estimation.

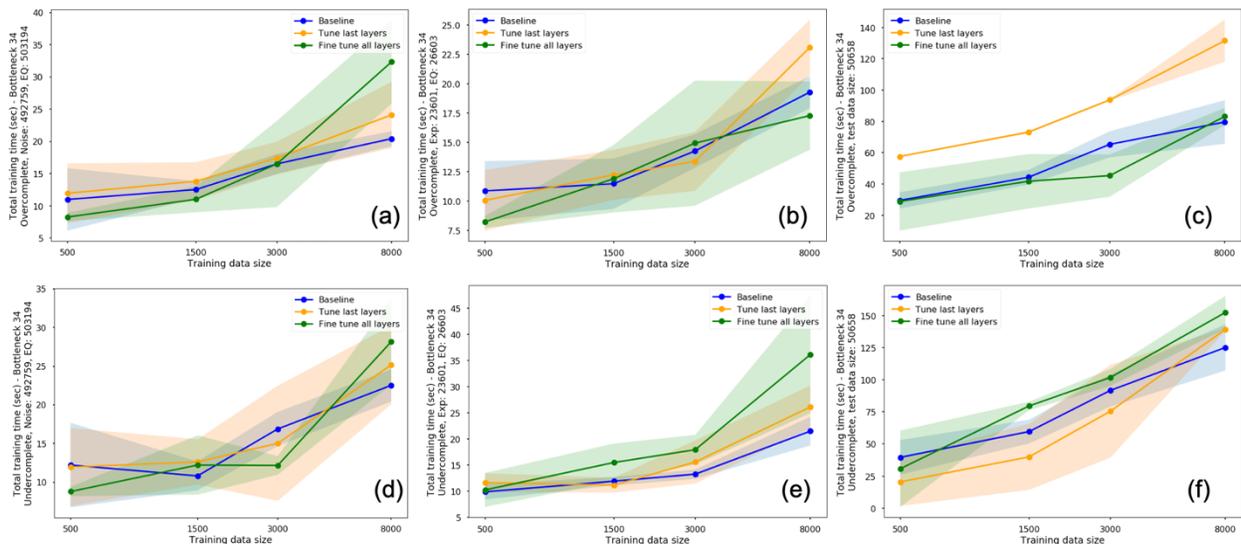

**Figure S16.** Zoomed in view of training time for different applications. See figure 8 for the whole view.

Data Distribution Figures

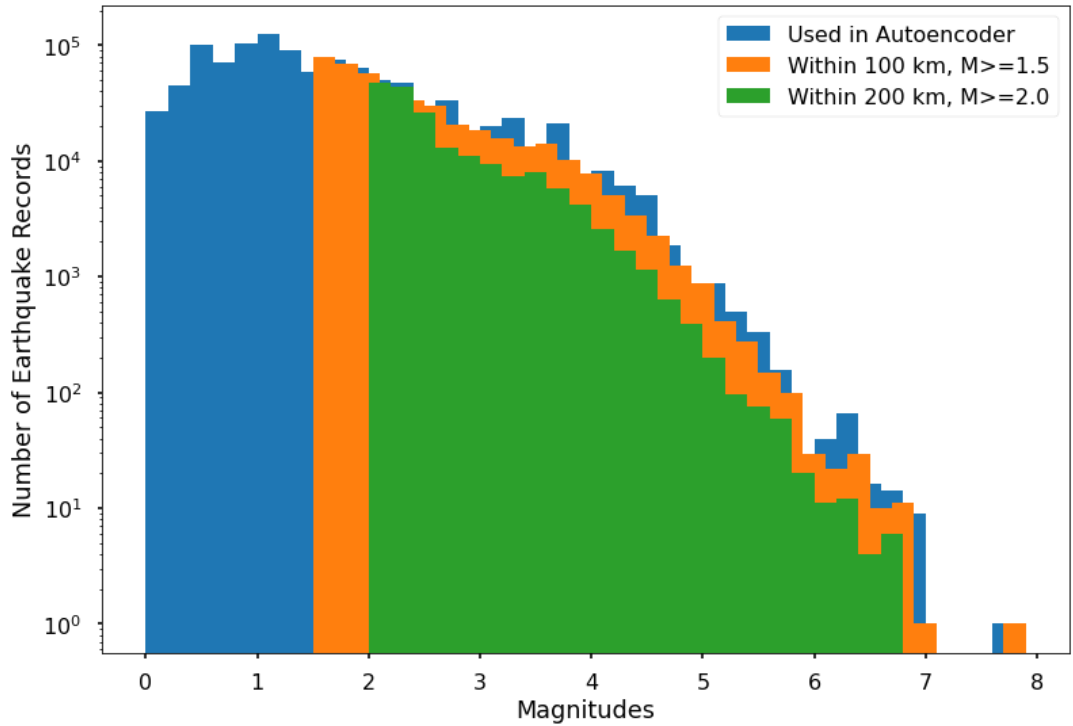

Figure S17. Magnitude distribution for the earthquake records in STEAD both used for the autoencoder and the phase picking (subset using distance and magnitude threshold)

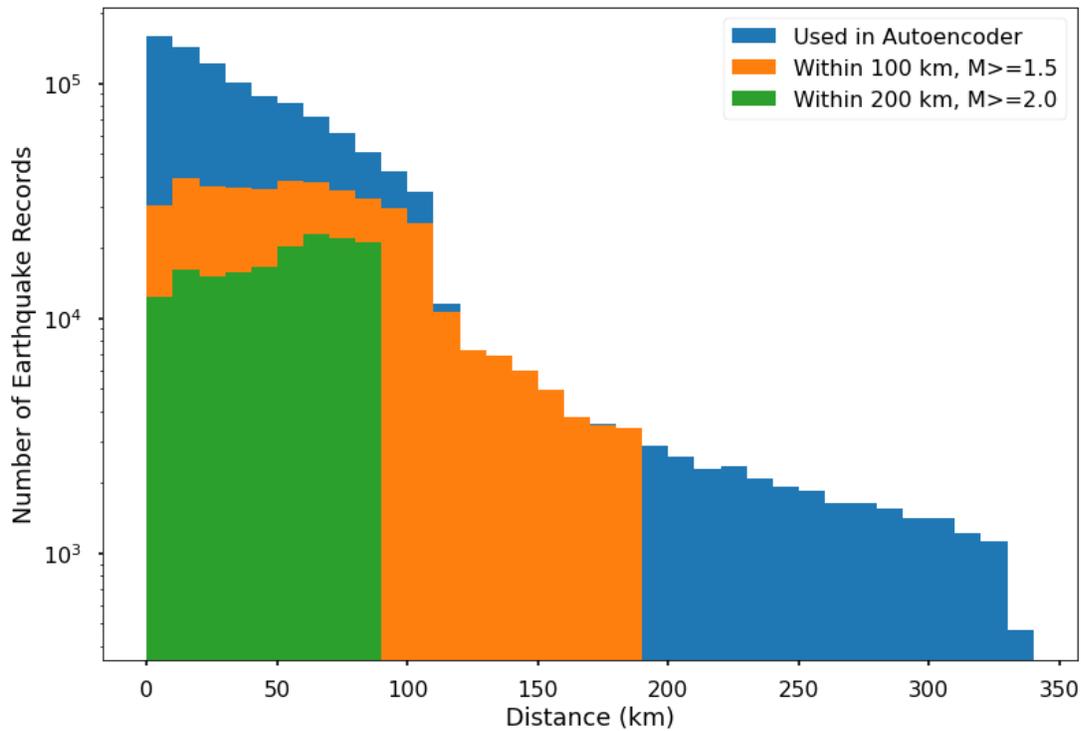

Figure S18. Distance distribution for the earthquake records in STEAD both used for the autoencoder and the phase picking (subset using distance and magnitude threshold)

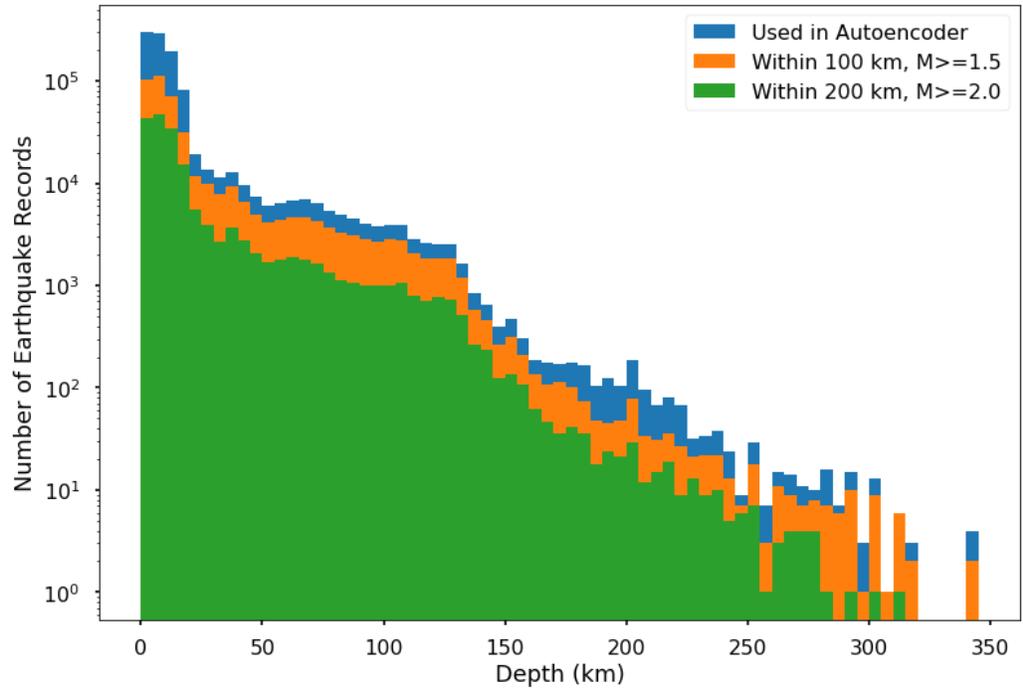

Figure S19. Depth distribution for the earthquake records in STEAD both used for the autoencoder and the phase picking (subset using distance and magnitude threshold)

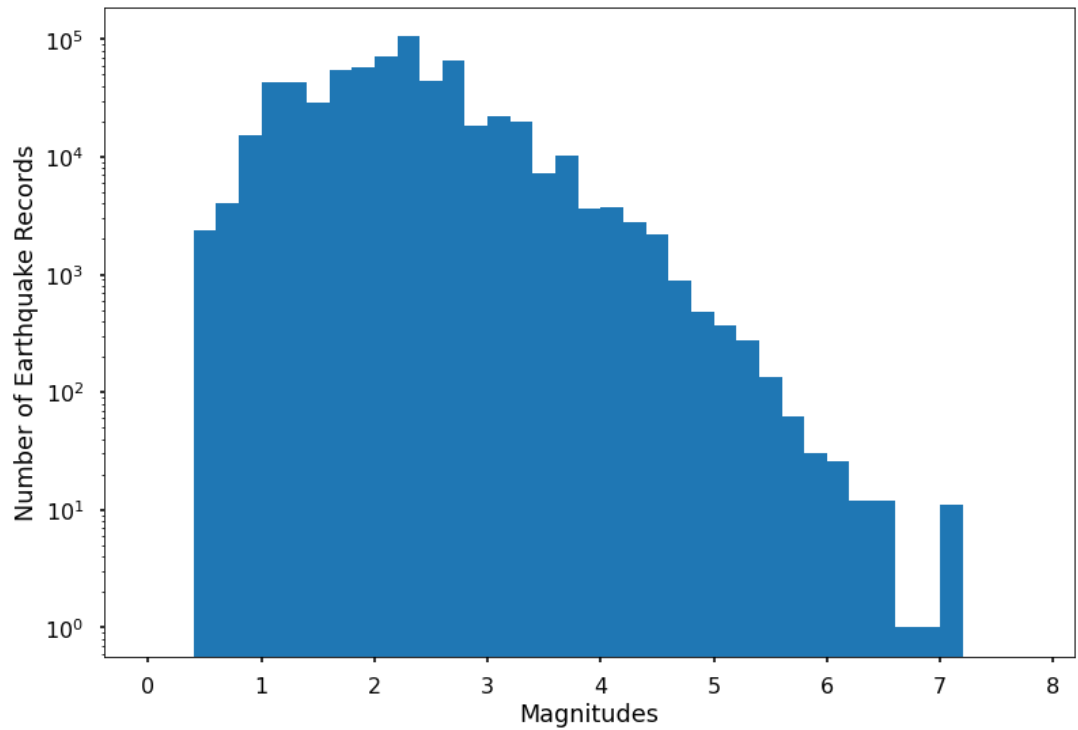

Figure S20. Magnitude distribution for the earthquake records in LEN_DB

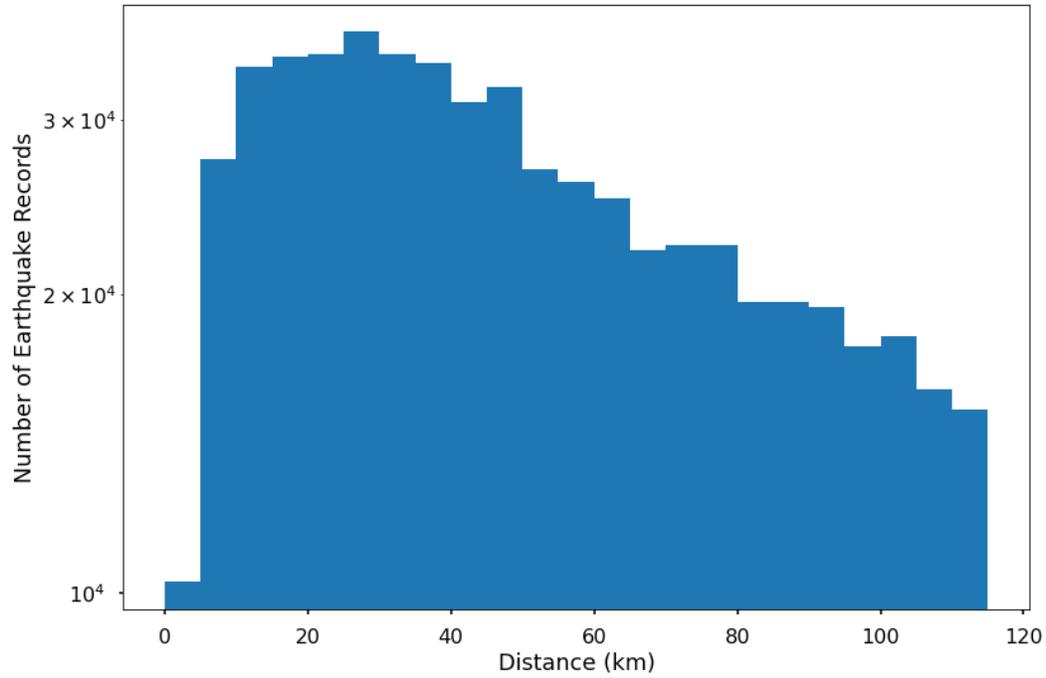

Figure S21. Event and receiver distance distribution for the earthquake records in LEN_DB

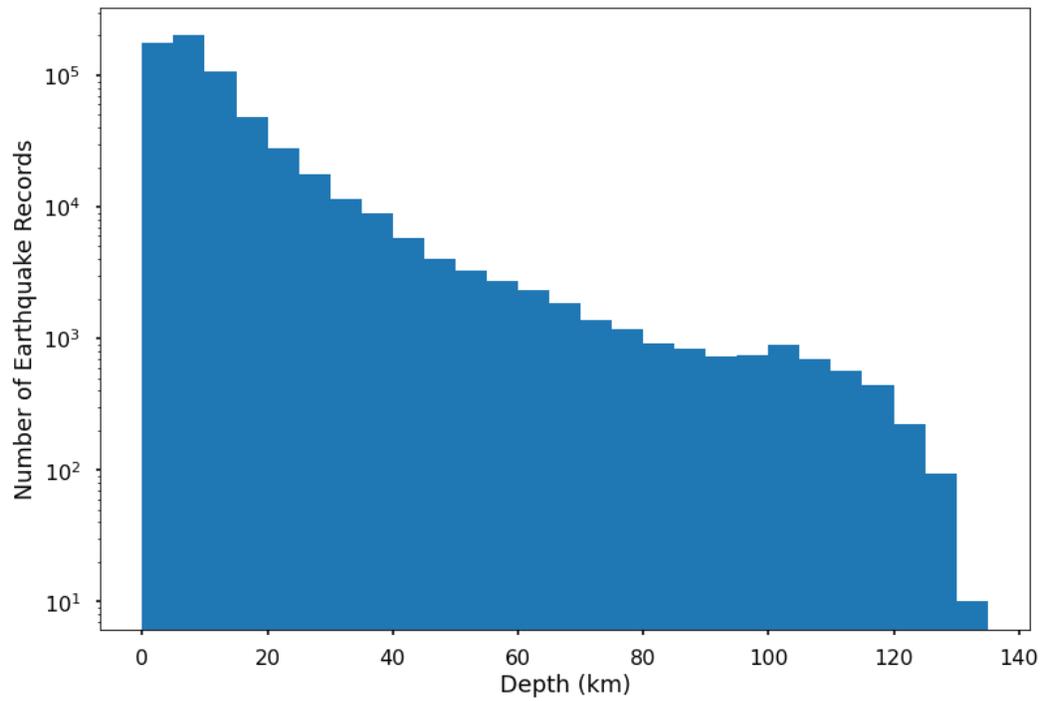

Figure S22. Depth distribution for the earthquake records in LEN_DB

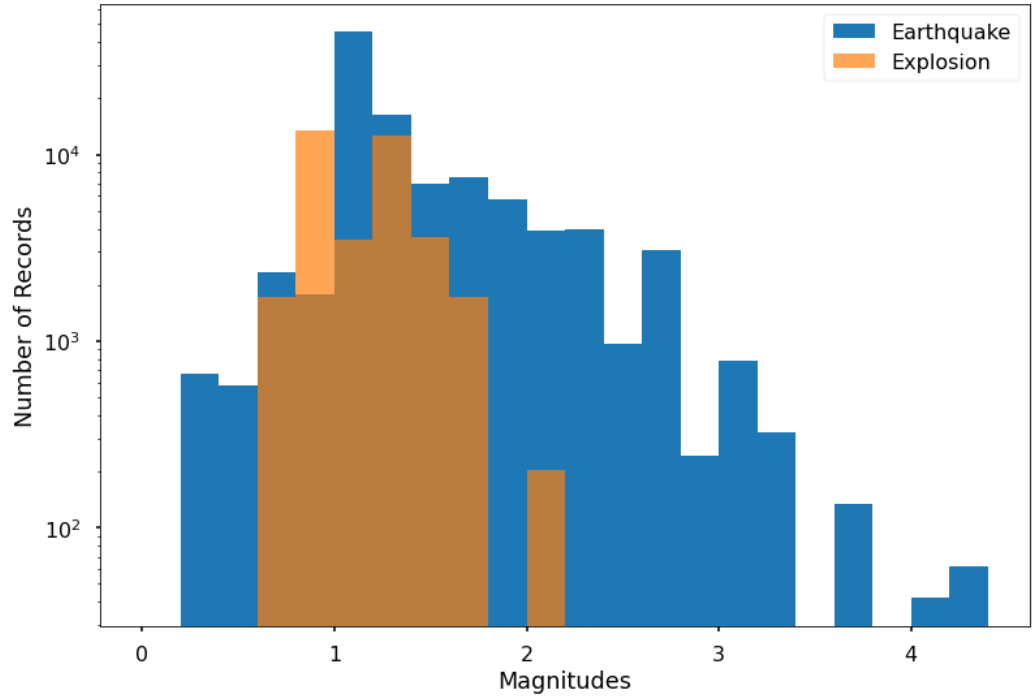

Figure S23. Magnitude distribution for the earthquake and explosion records in the explosion vs. earthquake problems.

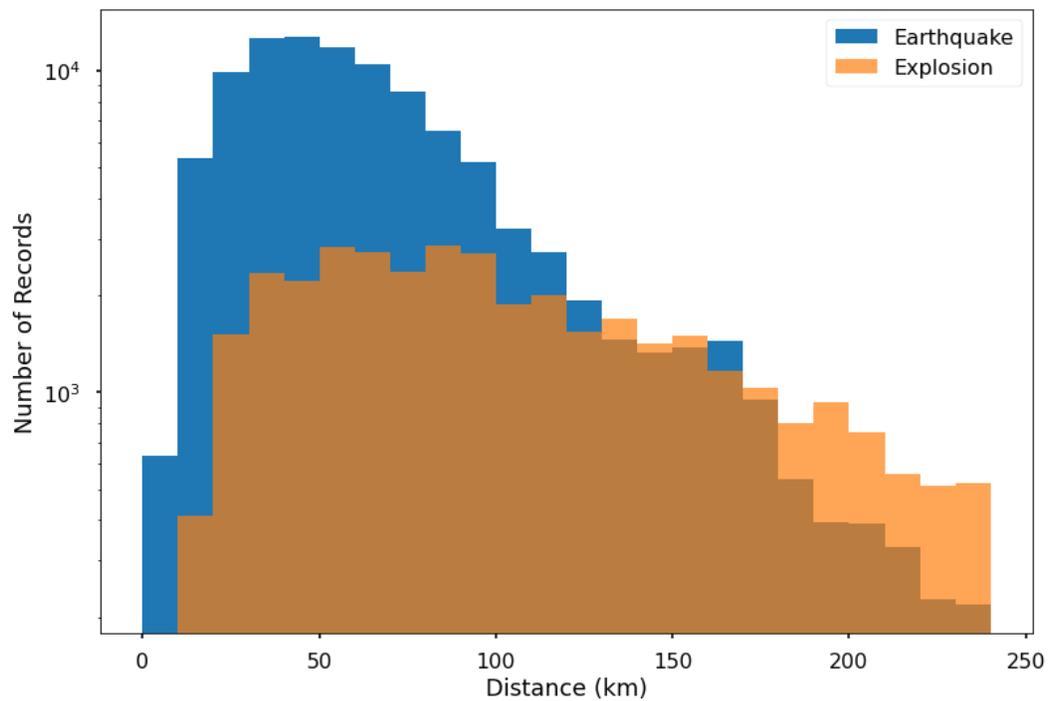

Figure S24. Event and receiver distance distribution for the earthquake and explosion records in the explosion vs. earthquake problems.

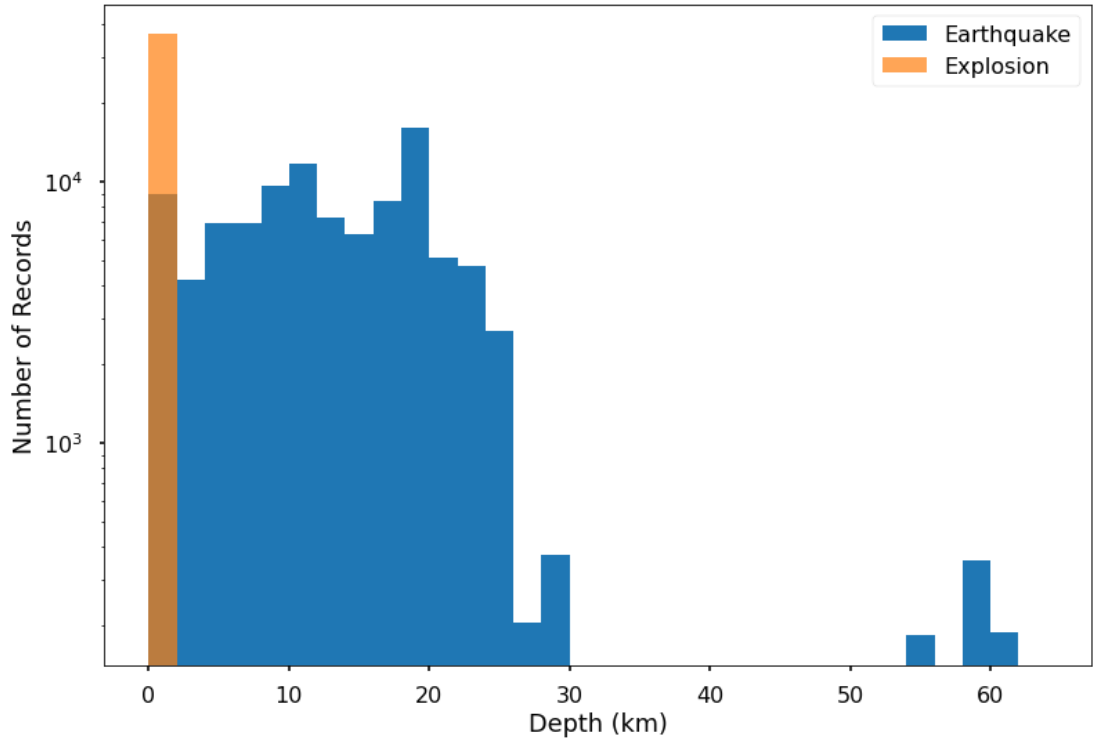

Figure S25. Depth distribution for the earthquake records in LEN_DB

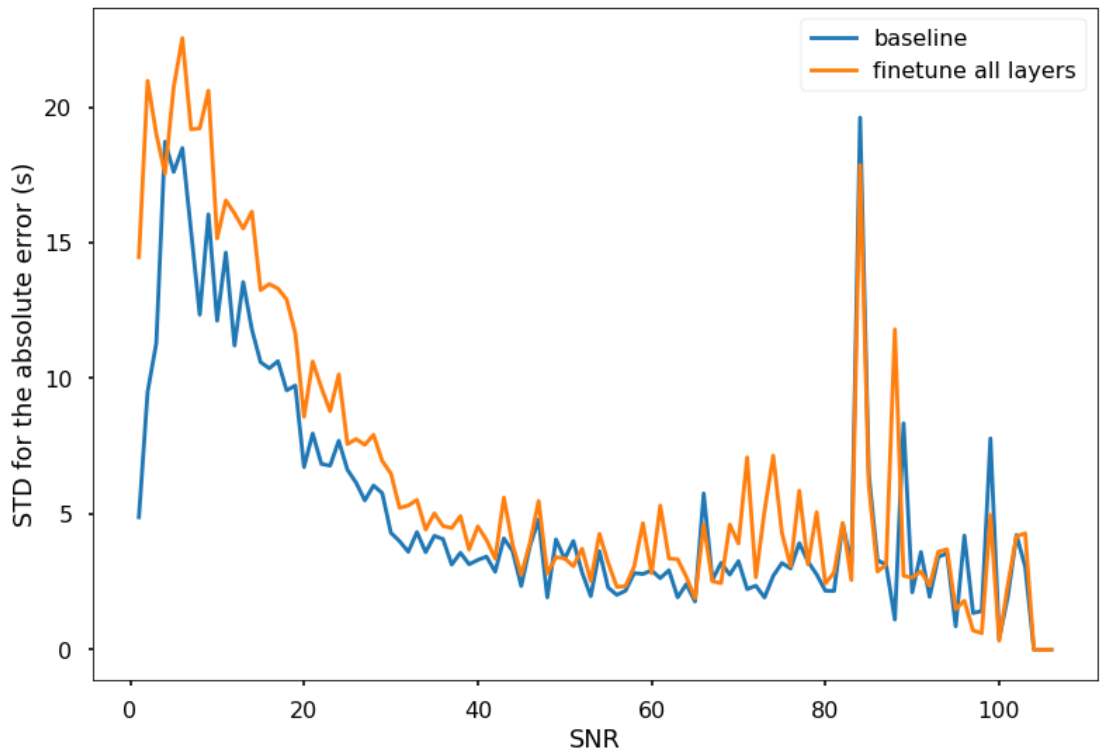

Figure S26. Standard Deviation (STD) of the phase picking absolute errors versus the signal noise ratio (SNR) on the test data when training data is 300,000 for the baseline model and the model with fine tuning all layers.

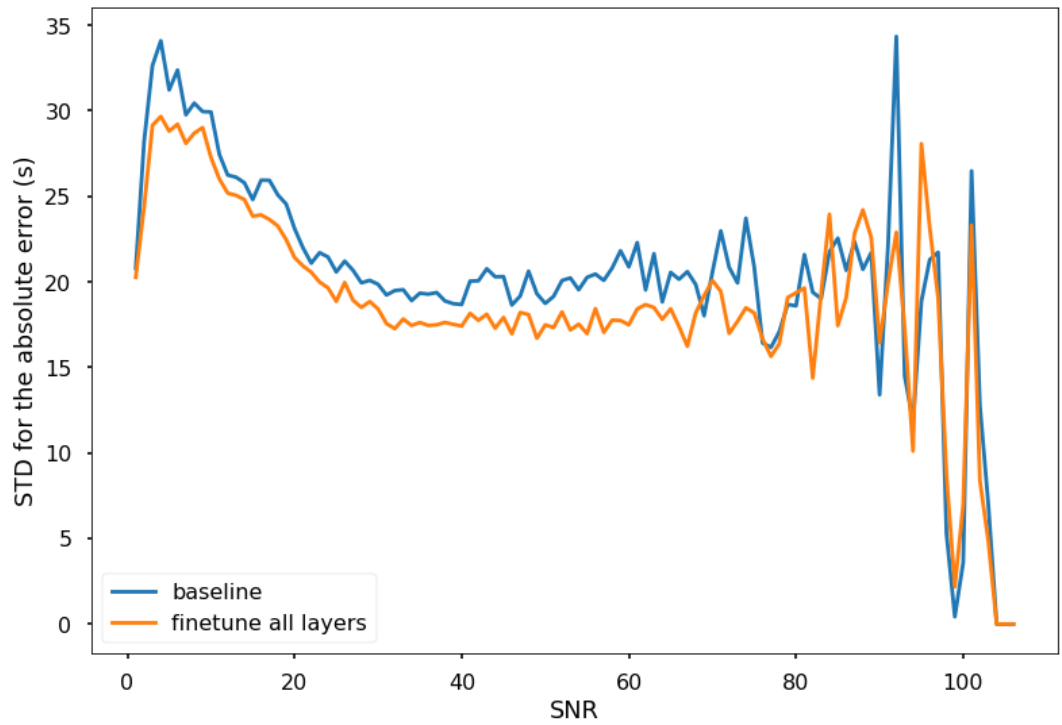

Figure S27. Standard Deviation (STD) of the phase picking absolute errors versus the signal noise ratio (SNR) on the test data when training data is 500 for the baseline model and the model with fine tuning all layers.